\documentclass[10pt,conference]{IEEEtran}

\usepackage{epsfig}
\usepackage{times}
\usepackage{latexsym}
\usepackage{subfigure}
\usepackage{bbold,bbm,amsfonts,amsmath}
\usepackage{graphicx}

\newtheorem{theorem}{Theorem}[section]

\newtheorem{definition}{Definition}[section]

\newtheorem{discussion}{Discussion}[section] 

\newcommand{\remove}[1]{}

\newcommand{\notes}[1]{}

\newcommand{\bmath}[1]{\mbox{\boldmath$#1$}}
\newcommand{\first}[1]{$1^{\mathrm{st}}$}
\newcommand{\second}[1]{$2^{\mathrm{nd}}$}

\begin{document}

\title{Modeling Multi-Cell IEEE 802.11 WLANs \\ with Application to
  Channel Assignment} 

\author{\IEEEauthorblockN{Manoj K. Panda and Anurag Kumar} 
\IEEEauthorblockA{Department of Electrical Communication Engineering
  \\ Indian Institute of Science, Bangalore -- 560012. \\ Email:
  \{manoj,anurag\}@ece.iisc.ernet.in} 
}

\maketitle

\begin{abstract}
We provide a simple and accurate analytical model for multi-cell
\textit{infrastructure} IEEE 802.11 WLANs. Our model applies if the
\textit{cell radius}, $R$, is much smaller than the \textit{carrier
  sensing range}, $R_{cs}$. We argue that, the condition $R_{cs} >> R$
is likely to hold in a dense deployment of Access Points (APs) where,
for every client or station (STA), there is an AP very close to the
STA such that the STA can associate with the AP at a high physical
rate. We develop a scalable \textit{cell level} model for such WLANs
with saturated AP and STA queues as well as for TCP-controlled long
file downloads. The accuracy of our model is demonstrated by
comparison with \textit{ns-2} simulations. We also demonstrate how our
analytical model could be applied in conjunction with a Learning
Automata (LA) algorithm for optimal channel assignment. Based on the
insights provided by our analytical model, we propose a simple
decentralized algorithm which provides \textit{static} channel
assignments that are \textit{Nash equilibria in pure strategies} for
the objective of maximizing \textit{normalized network
  throughput}. Our channel assignment algorithm requires neither any
explicit knowledge of the topology nor any message passing, and
provides assignments in only as many steps as there are channels. In
contrast to prior work, our approach to channel assignment is based on
the \textit{throughput metric}. 
\end{abstract}


\section{Introduction}
\label{sec:introduction}

\begin{figure}[tb]
\centering \
  \begin{minipage}{8cm}
  \begin{center}
\psfig{figure=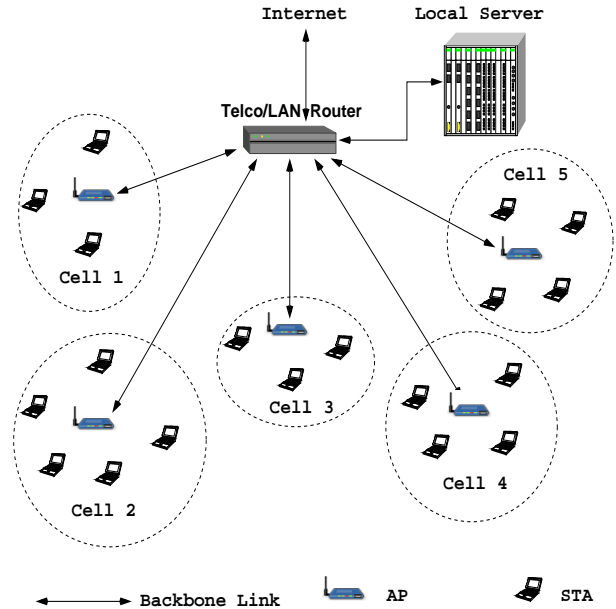,height=8cm,width=8cm}
     \caption{A multi-cell infrastructure WLAN: DCF is used only for
       communications within the cells. Connectivity to nodes outside
       the cell is provided over a separate backbone which connects
       the APs to the Internet through a local server and a LAN
       router. \label{fig:multicell-infrastructure-wlan}} 
  \end{center}
  \end{minipage}
\end{figure}

This paper is concerned with \textit{infrastructure mode} Wireless
Local Area Networks (WLANs) that use the Distributed Coordination
Function (DCF) Medium Access Control (MAC) protocol as defined in the
IEEE 802.11 standard \cite{wanet.IEEE8021199standard}. Such WLANs
contain a number of Access Points (APs). Every client station (STA) in
the WLAN associates with exactly one AP. Each AP, along with its
associated STAs, defines a \textit{cell}. Each cell operates on a
specific channel. Cells that operate on the same channel are called
\textit{co-channel}. Thus, in our setting, DCF is used only for
\textit{single-hop} communication within the cells, and STAs can
access the Internet only through their respective APs, which are
connected to the Internet by a high-speed wire line local area
network. Figure \ref{fig:multicell-infrastructure-wlan} depicts such a
\textit{multi-cell} infrastructure WLAN.

To support the ever-increasing user population at high access speeds,
WLANs are resorting to dense deployments of APs where, for every STA,
there exists an AP close to the STA with which the STA can associate
at a high Physical (PHY) rate
\cite{wanet.murty_etal08dendeAP}. However, as the density of APs
increases, cell sizes become smaller and, since the number of
non-overlapping channels is limited\footnote{For example, the number
  of non-overlapping channels in 802.11b/g is 3 and that in 802.11a is
  12.}, co-channel cells become closer. Nodes in two closely located
co-channel cells can suppress each other's transmissions via carrier
sensing and interfere with each other's receptions causing packet
losses. Thus, capacity might sometimes degrade with increased AP
density \cite{wanet.ergin08exptIntercellInterference},
\cite{wanet.belanger07novarumReport}. Clearly, effective planning and
management are essential for achieving the benefits of dense
deployments of APs. It has been demonstrated that a dense deployment
of APs, along with careful channel assignment and user association
control, can enhance the capacity by as much as 800\%
\cite{wanet.murty_etal08dendeAP}; but the technique has been tested on
a small scale\footnote{In \cite{wanet.murty_etal08dendeAP}, to provide
  connectivity at high PHY rates, 24 APs were deployed in an area
  which could have been covered by a single AP at a low PHY
  rate.}. Large-scale WLANs are difficult to plan and manage since
good network engineering models are lacking. 

Much of the earlier work on modeling WLANs deals with single-AP
networks or the so-called \textit{single cells}
\cite{wanet.bianchi00performance},
\cite{wanet.cali-etal00throughput-limit},
\cite{wanet.kumar_etal07new_insights}. The existing performance
analyses of multi-cell WLANs (i.e., WLANs consisting of multiple APs)
are mostly based either on simulations or small-/medium-scale
experiments \cite{wanet.murty_etal08dendeAP},
\cite{wanet.ergin08exptIntercellInterference},
\cite{wanet.belanger07novarumReport}. Studying large-scale WLANs by
simulations and experiments is both expensive and time-consuming. It
is, therefore, important to develop an analytical understanding of
such WLANs in order to derive insights into the system dynamics. The
insights thus obtained can be applied: (i) to facilitate planning and
management, and (ii) to develop efficient adaptive schemes for making
the WLANs \textit{self-organizing} and \textit{self-managing}. In this
paper, we first develop an analytical model for 802.11-based
multi-cell WLANs and then apply our model to the task of channel
assignment. \\

\noindent \textbf{Our Contributions:} We make the following
contributions: 

\begin{itemize}

\item We identify a condition, which we call the \textit{Pairwise
  Binary Dependence} (PBD) condition, under which multi-cell WLANs can
  be modeled at the \textit{cell level} (see A.\ref{item:AP-STA-close}
  in Section \ref{sec:model-assumptions}). 

\item We develop a scalable cell level model for multi-cell WLANs with
  \textit{arbitrary cell topologies} (Section
  \ref{sec:analysis-multicell-arbitrary-cell-topology}). 

\item We extend the single cell TCP analysis of
  \cite{wanet.bruno08TCPeqvSatModel} to multiple interfering cells
  (Section \ref{subsec:TCP-traffic}). 

\item We demonstrate how our model could be applied in conjunction
  with a Learning Automata (LA) algorithm for optimal channel
  assignment (Section \ref{sec:channel-assignment-algorithm}). 

\item Based on the insights provided by our analytical model, we
  propose a simple decentralized algorithm which can provide fast
  \textit{static} channel assignments (Section
  \ref{sec:simple-algorithm}). \\ 

\end{itemize}

Unlike the \textit{node level} models \cite{wanet.boorstyn87multihop},
\cite{wanet.garetto_etal08starvation} or the \textit{link level}
models \cite{wanet.wang-kar05multihop},
\cite{wanet.durvy08bordereffect} reported earlier in the
context of \textit{multi-hop ad hoc} networks, our cell level model
does not require modeling the activities of every single node
(resp. link). Thus, the complexity of our cell level model increases
with the number of cells rather than the number of nodes
(resp. links). However, \textit{our model can account for the number
  of nodes in each cell which may differ across the cells}. 

We argue that the PBD condition is likely to hold in a dense
deployments of APs (see Sections \ref{sec:motivation-contribution} and
\ref{subsec:discussion-assumption4}). Hence, our cell level model can
be applied to obtain a \textit{first-cut} understanding of large-scale
WLANs. 

Our cell level model is based on the channel contention model of
Boorstyn et al. \cite{wanet.boorstyn87multihop} and the transmission
attempt model of \cite{wanet.kumar_etal07new_insights}. Thus, our
approach is similar to that of
\cite{wanet.garetto_etal08starvation}. However, our model of
transmission attempt process is simpler (see the discussion following
Equation \ref{eqn:lambda_i-multicell-CTMC}) and the closed-form
expressions for collision probabilities and cell throughputs that we
derive are new. We also provide new insights. 

Our channel assignment algorithm provides assignments that are
\textit{Nash equilibria in pure strategies} for the \textit{normalized
  network throughput maximization} objective (see Section
\ref{sec:simple-algorithm}). Furthermore it provides an assignment in
only $M$ steps where $M$ denotes the number of available channels. In
contrast to prior work, our approach to channel assignment is based on
the \textit{throughput metric} (see the last paragraph in Section
\ref{sec:relevant-works}). Our channel assignment algorithm is
motivated by our analytical model which requires the knowledge of the
topology of APs. However, our channel assignment algorithm requires
neither any topology information nor any message passing. To be
specific, our channel assignment algorithm takes the help of the
network to discover an optimal assignment (in the sense mentioned
above) in a completely decentralized manner. 

The remainder of this paper is organized as follows. In Section
\ref{sec:relevant-works}, we discuss the related literature. In
Section \ref{sec:motivation-contribution}, we provide the motivation
for our simple cell level model. In Section
\ref{sec:model-assumptions}, we provide our network model and
summarize our key modeling assumptions. The analytical model is
developed in Section
\ref{sec:analysis-multicell-arbitrary-cell-topology}. In Section
\ref{sec:results}, we validate our model by comparing with
\textit{ns-2} simulations. In Section \ref{sec:design-example}, we
apply our model to compare three different channel assignments for a
12-cell network and emphasize the impact of carrier sensing in 802.11
networks. In Section \ref{sec:channel-assignment-algorithm}, we
demonstrate how our analytical model could be applied along with a
Learning Automata (LA) algorithm for optimal channel assignment. We
propose a simple and fast decentralized channel assignment algorithm
in Section \ref{sec:simple-algorithm}, and summarize the conclusions
in Section \ref{sec:conclusion}. Two lengthy derivations have been
deferred to the appendices at the end of the paper.

\section{Related Literature}
\label{sec:relevant-works}

Modeling of \textit{multi-hop ad hoc networks} is closely related to
that of multi-cell WLANs. In both cases, the \textit{hidden node} and
the \textit{exposed node} problems \cite{wanet.roy06PCS} are known to
be the two main capacity-degrading factors. Multi-hop ad hoc networks
and multi-cell WLANs are much harder to model than single cells
precisely because of the presence of hidden and exposed nodes. Since
the evolution of activities of each node is, in general, statistically
different from that of the others', one needs to model the system at
the \textit{node level} \cite{wanet.boorstyn87multihop},
\cite{wanet.garetto_etal08starvation}, or at the \textit{link level}
\cite{wanet.wang-kar05multihop}, \cite{wanet.durvy08bordereffect},
i.e., the activities of every single node or link and the interactions
among them need to be modeled. 

In the context of CSMA based multi-hop packet radio networks, Boorstyn
et al. \cite{wanet.boorstyn87multihop} proposed a Markovian model with
Poisson packet arrivals and arbitrary packet length
distributions. Wang and Kar \cite{wanet.wang-kar05multihop} adopted
the node level Boorstyn model to develop a link level model for 802.11
networks and studied the fairness issues. As a simplification, they
assumed a fixed \textit{contention window}. Garetto et
al. \cite{wanet.garetto_etal08starvation} extended the Boorstyn model
to multi-hop 802.11 networks. They computed the packet loss
probabilities and the throughputs per node accounting for the details
of the 802.11 protocol. In particular, they incorporated the evolution
of contention windows, which was missing in
\cite{wanet.wang-kar05multihop}, by applying the analysis of
\cite{wanet.kumar_etal07new_insights}. However, a direct application
of the Boorstyn model required two different activation rates, namely,
$\lambda$ and $g$, which complicated their model (see the discussion
following Equation \ref{eqn:lambda_i-multicell-CTMC}). We develop a
simpler model by slightly modifying the Boorstyn contention model and
making it particularly suitable for 802.11 networks. 

In the context of multi-cell infrastructure WLANs, Nguyen et
al. \cite{wanet.nguyen07stochGeom} proposed a model for dense
802.11 networks. To keep their interference analysis tractable, they
assumed all the APs in the WLAN to be operating on the same
channel. Recently, Bonald et
al. \cite{wanet.bonald08multicellprocsharing} applied the concept of
``exclusion region'' to model a multi-AP WLAN as a network of
multi-class processor-sharing queues with state-dependent service
rates. The concept of ``exclusion domains'' has also been applied in
\cite{wanet.durvy08bordereffect} to study the long-term fairness
properties of large networks. The concept of exclusion says that,
among a set of links that are in conflict, at most one can be active
at any point of time. As argued in \cite{wanet.durvy08bordereffect}
and \cite{wanet.bonald08multicellprocsharing}, exclusion can be
enforced in 802.11 networks by the RTS/CTS mechanism. However,
suppression of interferers by RTS/CTS is not perfect because: (1) RTS
frames can also collide and they must be retransmitted to enforce
exclusion, (2) the overheads due to RTS/CTS that are transmitted at
low rates and the capacity wastage due to RTS collisions might be
significant especially when data payloads are transmitted at high
rates, and most importantly, (3) RTS/CTS cannot completely eliminate
hidden node collisions since nodes that cannot decode the CTS frames
may collide with DATA frames that are longer than the Extended Inter
Frame Space (EIFS). Thus, \textit{``exclusion'' is strongly dependent
  on the virtual carrier sensing enforced by the RTS/CTS mechanism and
  is, essentially, a modeling approximation which ignores the
  possibility of hidden node collisions}. The ineffectiveness of
RTS/CTS is well known \cite{wanet.xuGerla-etal02RTS-CTS-effectiveness}
and Basic Access can outperform RTS/CTS at high data rates even when
hidden nodes are present
\cite{wanet.tinnirello-etal05RTS-CTS-effectiveness}. 



Channel assignment in WLANs has been extensively studied (see, e.g.,
\cite{wanet.leung_kim03_freq_assignment},
\cite{FAP.mishra06clientdriven}, \cite{FAP.mishra06Maxchop},
\cite{wanet.leith_clifford06CFL},
\cite{wanet.kauffmann07selfOrganization},
\cite{wanet.villegas_etal08WiComMobCom} and the references
therein). Much of the existing work on channel assignment proposes to
minimize the global interference power or maximize the global Signal
to Noise and Interference Ratio (SINR) without taking into account the
combined effect of the PHY and the MAC layers. Due to carrier sensing,
nodes in 802.11 networks get opportunity to transmit for only a
fraction of time, and this must be accounted for when computing the
global interference power or SINR. Such an approach is found only in
\cite{wanet.leung_kim03_freq_assignment} where the authors propose to
maximize a quantity called ``effective channel utilization''. In
reality, however, end users are more interested in the ``throughputs''
and, ideally, the objective should be to maximize the \textit{sum
  utility} of throughputs. A heuristic method for computing
throughputs can be found in
\cite{wanet.villegas_etal08WiComMobCom} where the authors assume long
term fairness among the nodes that are mutually interfering. The
heuristic adopted in \cite{wanet.villegas_etal08WiComMobCom} says
that, ignoring collisions and channel errors, a node obtains
approximately $\frac{1}{n}$ of the maximum capacity where $n$ denotes
the cardinality of the maximum \textit{clique}\footnote{A clique is a
  set of mutually interfering nodes.} to which the node belongs. In
reality, the size of the maximum clique to which a node belongs can be
very different from the size of the maximum clique to which one of its
interfering neighbor belongs, and this leads to inconsistent
predictions for node throughputs. Nevertheless, the authors in
\cite{wanet.villegas_etal08WiComMobCom} claim that, their heuristic
can predict the \textit{global} throughput quite well even though
individual node throughputs may not be. We remark that, the multi-cell
model in \cite{wanet.bonald08multicellprocsharing} does not actually
model the DCF contention and adopts a model of capacity sharing
similar to the heuristic of \cite{wanet.villegas_etal08WiComMobCom}. 

In summary, due to the inherent complexity of the problem, simplifying
assumptions have often been made to develop tractable analytical
model. The common simplification is to ignore hidden node collisions
through suitable assumptions. We also ignore hidden node collisions to
develop our model. However, unlike
\cite{wanet.bonald08multicellprocsharing}, our model can be applied
either with the Basic Access mechanism or with the RTS/CTS mechanism
provided that the PBD condition holds (see A.\ref{item:AP-STA-close}
in Section \ref{sec:model-assumptions}). To our knowledge, an accurate
throughput based approach which accounts for PHY-MAC interactions does
not exist. Thus, we first develop a simple and accurate throughput
model for multi-cell WLANs and then apply our model to the task of
channel assignment.

\section{Motivation for a Cell Level Model}
\label{sec:motivation-contribution}

\begin{figure}[tb]
\centering \
  \begin{minipage}{8cm}
  \begin{center}
\psfig{figure=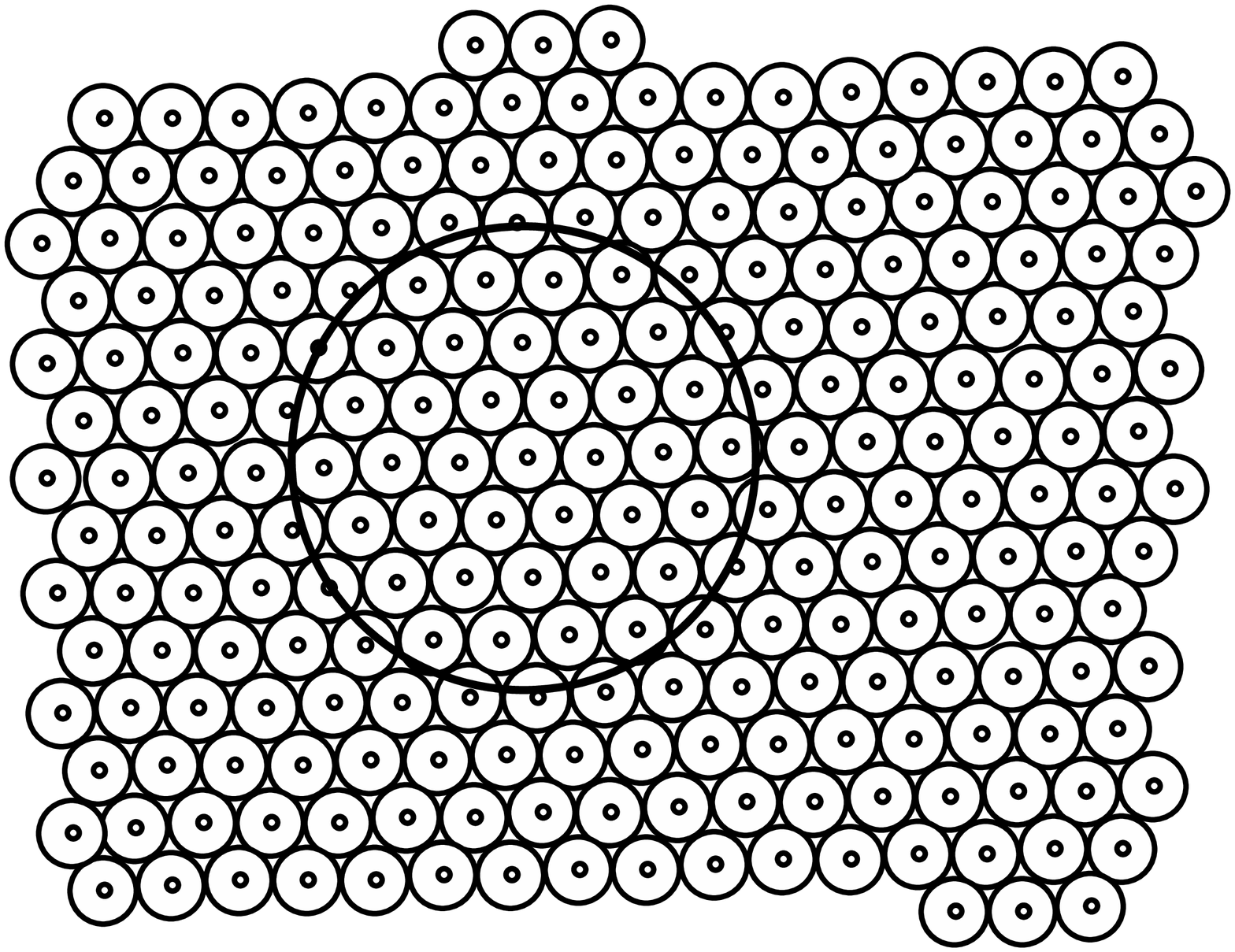,height=6cm,width=8cm}
     \caption{A dense AP network: Small circular areas represent cells
       with radius $R$. APs are shown as tiny circles at the centers of
       the cells. The big circular disk represents the area covered by
       the carrier sensing range $R_{cs}$ of one of the APs. Observe
       that, with $R_{cs} >> R$, other cells are either almost
       completely covered or almost not covered at all through carrier
       sensing. This indicates that, with $R_{cs} >> R$, nodes
       belonging to the same cell ``perceive'' identical medium
       activities, at least approximately. \label{fig:PBD-idea}} 
  \end{center}
  \end{minipage}
\end{figure}

In a dense deployment of APs with denser user population, it seems
practically impossible to apply a node or a link level model for
planning and managing the network. However, we can exploit a specific
characteristics of dense deployments. Let $R$ denote the \textit{cell
  radius}, i.e., $R$ is the maximum distance between an AP and the
STAs associated to it. Let $R_{cs}$ denote the \textit{carrier sensing
  range}, i.e., $R_{cs}$ is the distance up to which carrier sensing
is effective\footnote{Precise definition of carrier sensing range can
  be found in \cite{wanet.roy06PCS} and
  \cite{wanet.xu_vaidya_infocom05}.}. We observe that, ${R_{cs} >> R}$
is likely to hold in a dense deployment of APs where, for every STA,
there is an AP very close to the STA. With ${R_{cs} >> R}$, the
network model can be simplified in the following ways: 

\begin{enumerate}

\item Since any transmitter `T' is within a small distance $R$ from
  its receiver `R', a node `H' that is beyond a distance $R_{cs}$ from
  `T' (i.e., a \textit{hidden node}) is unlikely to interfere with
  `R'\footnote{Ignoring noise, we have, SINR $\geq
    \left(\frac{R_{cs}}{R}\right)^{\nu}$, where $\nu$ is called the
    \textit{path loss exponent} and takes a value between 2 to 4.},
  i.e., carrier sensing would avoid much of the co-channel
  interference and we may ignore collisions due to hidden
  nodes. 

\item If Node-1 in Cell-1 can sense the transmissions by Node-2 in
  Cell-2, then it is likely that all the nodes in Cell-1 can sense the
  transmissions from all the nodes in Cell-2 and vice versa, i.e., we
  may assume that \textit{nodes belonging to the same cell have an
    identical view of the rest of the network and interact with the
    rest of the network in an identical manner}. Figure
  \ref{fig:PBD-idea} summarizes this idea. 

\end{enumerate}

The assumption that the AP and all its associated STAs have an
identical view of the network has been applied in a dense AP setting
\cite{wanet.murty_etal08dendeAP} where the authors approximate STA
statistics by statistics collected at the APs for efficiently managing
their network. We adopt this idea of \cite{wanet.murty_etal08dendeAP}
to develop an analytical model. We identify the locations of the STAs
with the locations of their respective APs, and treat a cell as a
single entity, thus yielding a scalable \textit{cell level}
model. Unlike a node level model, the complexity of a cell level model
increases with the number of cells rather than number of nodes. A
simple cell level model is particularly suitable for the task of
channel assignment since channels are assigned to cells rather than to
nodes. The foregoing simplification is extremely useful at the
network planning stage when the locations of the STAs are not known
but the locations of the APs and the expected number of users per cell
might be known. Furthermore, since much of the traffic in today's
WLANs is downlink, i.e., from the APs to the STAs, a large fraction of
channel time is occupied by transmissions from the APs. It is then
reasonable to develop a model based only on the topology of the APs
and the expected number of users per cell assuming that the users are
located close to their respective APs. 

The appropriate locations for placing the APs can be decided using
standard RF tools and the \textit{physical} topology of the APs can be
determined. Given the physical topology of the APs and a channel
assignment, we can obtain the \textit{logical} topology, i.e., the
topology of the co-channel APs corresponding to every channel. Given a
logical topology, our analytical model can predict the cell
throughputs accounting for the expected number of users per cell. The
cell throughputs thus obtained can provide the \textit{goodness} of
the assignment to a channel assignment algorithm based on which the
algorithm can determine a \textit{better} channel assignment. Given
the logical topology with the new channel assignment, our model can
provide the modified feedback and this procedure can be repeated
several times to arrive at an optimal assignment. We explain this
approach in Section \ref{sec:channel-assignment-algorithm} where we
apply our model in conjunction with a Learning Automata (LA) based
algorithm for optimal channel assignment.

\section{Network Model and Assumptions}
\label{sec:model-assumptions}

We consider scenarios with $R_{cs} >> R$. Based on our earlier
discussion, we assume that simultaneous transmissions by nodes that
are farther than $R_{cs}$ from each other results in successful
receptions at their respective receivers. We also assume that
simultaneous transmissions by nodes that are within $R_{cs}$ always
lead to packet losses at their respective receivers, i.e., we ignore
the possibility of \textit{packet capture}\footnote{The difficulty in
  modeling packet capture lies in analytically computing the capture
  probabilities. Given the capture probabilities, packet capture can
  also be incorporated along the lines of
  \cite{wanet.ramaiyan_kumar05capture}.}. For simplicity, we assume
\textit{symmetry} among the nodes and of the propagation
environment. In particular, this means that (i) nodes use identical
protocol parameters (e.g., contention windows, retry limit etc.), and
(ii) if Node-1 can sense Node-2, then Node-2 can also sense Node-1. 

We say that two nodes are \textit{dependent} if they are within
$R_{cs}$; otherwise, the two nodes are said to be
\textit{independent}. Two cells are said to be independent if every
node in a cell is independent w.r.t. every node in the other cell;
otherwise, the two cells are said to be dependent. Two dependent cells
are said to be \textit{completely dependent} if every node in a cell
is dependent w.r.t. every node in the other cell. In this broad
setting, our key assumptions are the following:

\renewcommand{\labelenumi}{A.\theenumi}

\begin{enumerate}
\item \label{item:orthogonal-channels-only} Only non-overlapping
  channels are used. 


\item \label{item:closet-AP-forever} Associations of STAs with APs are
  static. This implies that the number of STAs in a cell is fixed.

\item \label{item:AP-STA-close} \textbf{Pairwise Binary Dependence
  (PBD):} Any pair of cells is either independent or completely
  dependent. 

\item \label{item:no-channel-error} The STAs are so close to their
  respective APs that packet losses due to channel errors are
  negligible. 

\item \label{item:EIFSeqDIFS} The EIFS deferral has been disabled in
  the sense that medium access always starts after a Distributed
  Inter-Frame Space (DIFS) deferral. 

\end{enumerate}

\renewcommand{\labelenumi}{\theenumi.}

\subsection{Discussion of the PBD Condition and Assumption A.\ref{item:EIFSeqDIFS}} 
\label{subsec:discussion-assumption4}

The PBD condition is crucial to the analytical model being developed
in this paper. It includes the possibility that two given co-channel
cells can be independent, i.e., two co-channel cells can be so far
apart that activities in one cell do not affect the activities in the
other cell. However, if two cells are dependent, the PBD condition
rules out the possibility that only a subset of nodes in one cell can
sense a subset of nodes in the other cell. We note that, in a dense
deployment of APs, due to small cell radius $R$ and $R_{cs} >> R$,
complete dependence would hold up to $R_{cs}$ and independence would
hold beyond $R_{cs}$, i.e., the PBD condition would hold, at least
approximately. The PBD condition is a geometric property that enables
modeling at the cell level since, \textit{if the PBD condition holds,
  the relative locations of the nodes within a cell do not
  matter}. Also, any interferer `I', which must be within $R_{cs}$ of
a receiver `R', will also be within $R_{cs}$ of the transmitter `T' if
the PBD condition holds. Hence, either (i) `I' can start transmitting
at the same time as `T' causing \textit{synchronous} collisions at
`R', or (ii) `I' gets suppressed by T's RTS (or DATA) transmission
followed by R's CTS (or ACK) transmission since CTS and ACK frames are
given higher priority through SIFS ($<$ DIFS $<$ EIFS). Thus, if the
PBD condition holds, nodes do not require deferral by EIFS; deferral
by DIFS would suffice. Thus, we assume that contention for medium
access always begins after deferral by DIFS and we do not model the
impact of EIFS. We remark that certain commercial products, e.g.,
Atheros cards, have user-configurable EIFS and EIFS can be easily
disabled in such devices. 


\section{Analysis of Multi-Cell WLANs with Arbitrary Cell Topology}
\label{sec:analysis-multicell-arbitrary-cell-topology}

\begin{figure}[tb]
\centering \
  \begin{minipage}{3.75cm}
  \begin{center}
\includegraphics[height=1.25cm,width=3.75cm]{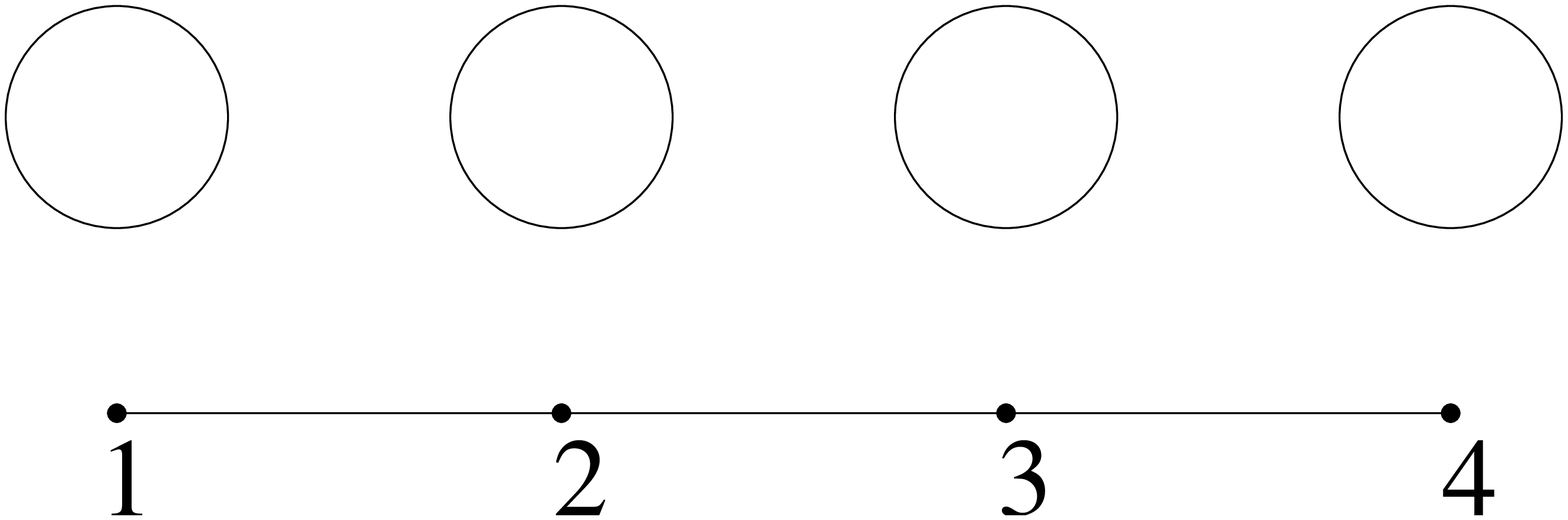}
\subfigure[]{\label{fig:fourCellLinear}}
  \end{center}
  \end{minipage}
\hfill
  \begin{minipage}{4cm}
  \begin{center}
\includegraphics[height=1.2cm,width=4cm]{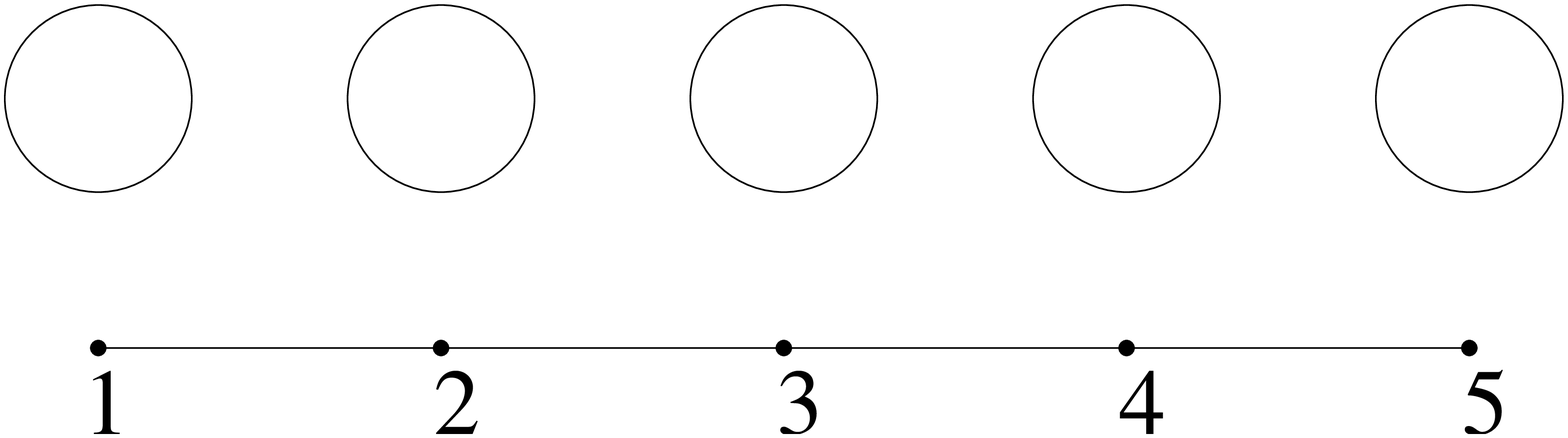}
\subfigure[]{\label{fig:fiveCellLinear}}
  \end{center}
  \end{minipage}
\hfill
  \begin{minipage}{4.25cm}
  \begin{center}
\includegraphics[height=2.4cm,width=4.25cm]{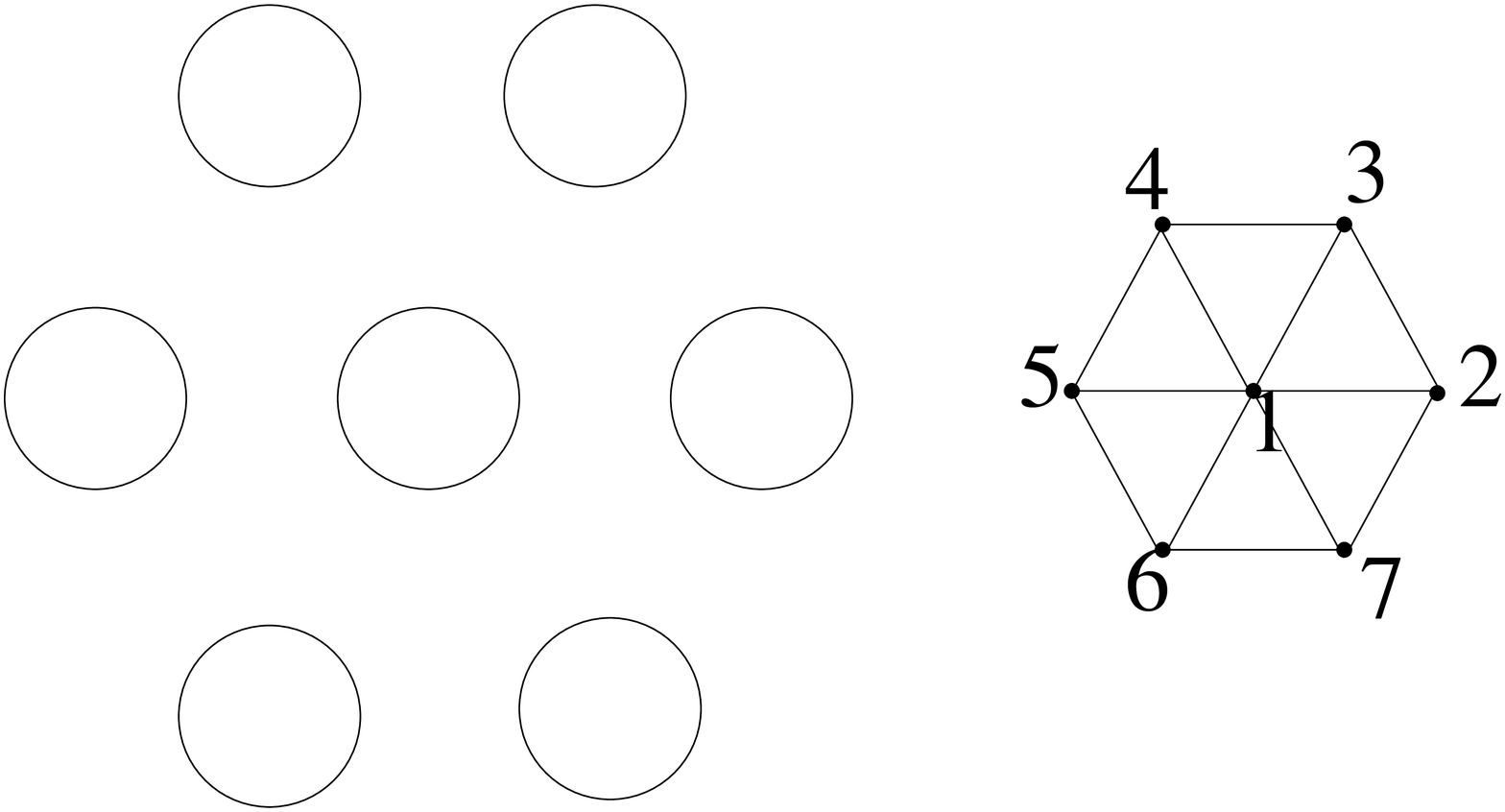}
\subfigure[]{\label{fig:sevenCellHexagonal}}
  \end{center}
 \end{minipage}
\hfill
  \begin{minipage}{3.5cm}
  \begin{center}
\includegraphics[height=3.25cm,width=3.5cm]{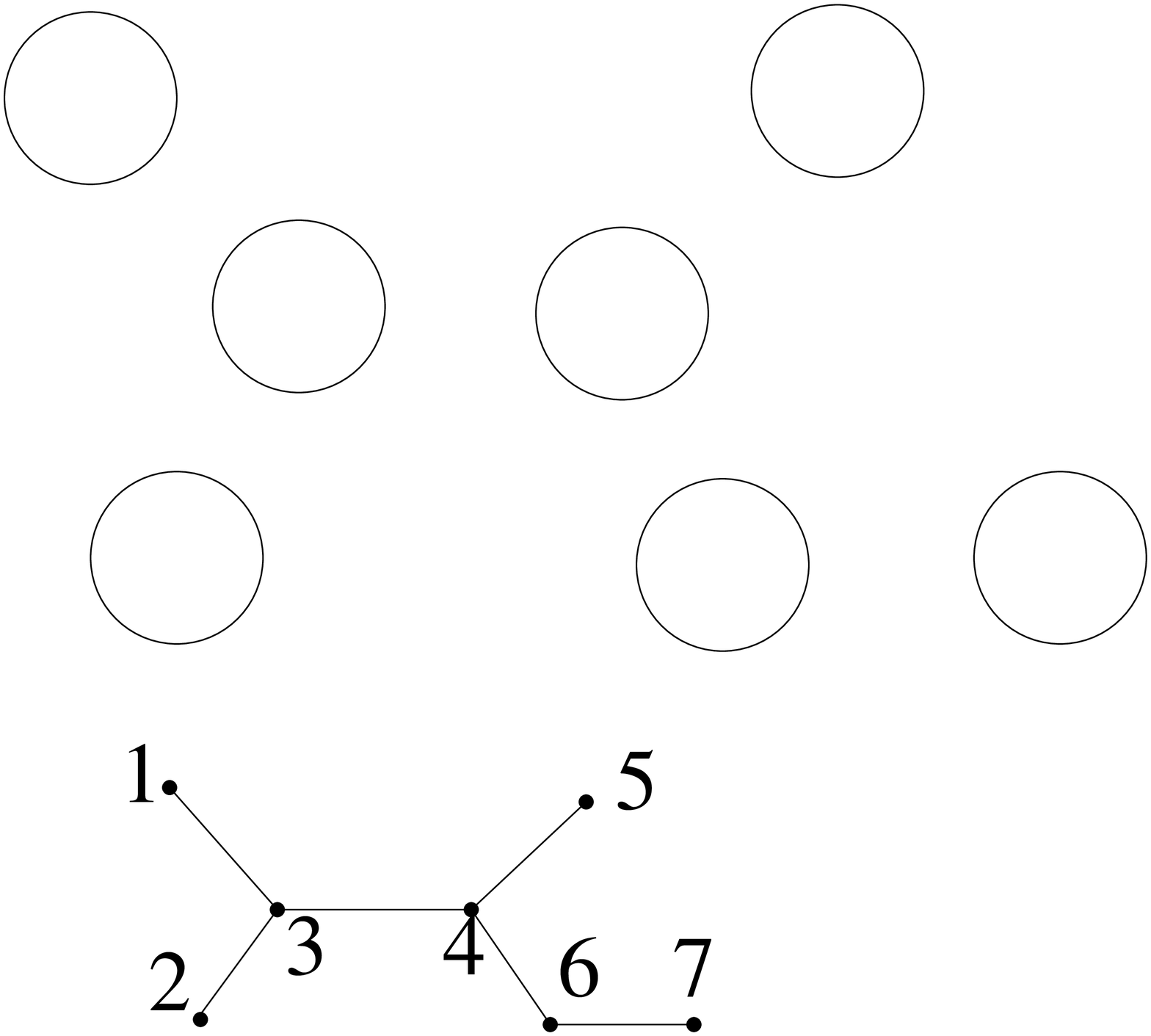}
\subfigure[]{\label{fig:sevenCellArbitrary}}
  \end{center}
 \end{minipage}
\hfill
\caption{Examples of multi-cell systems: (a) four linearly placed
  co-channel cells, (b) five linearly placed co-channel cells, (c)
  seven hexagonally placed co-channel cells, and (d) seven co-channel
  cells with an arbitrary cell topology. The cell level contention
  graphs have also been shown where the dots represent the
  cells. Neighbors have been joined by edges. For example (a), the
  pairs $\{1,2\}$, $\{2,3\}$ and $\{3,4\}$ are dependent and the pairs
  $\{1,3\}$, $\{1,4\}$ and $\{2,4\}$ are independent.} 
\end{figure}

In this section, we develop a cell level model for WLANs that satisfy
the PBD condition. We provide a generic model and demonstrate the
accuracy of our model by comparing with simulations of specific cell
topologies pertaining to: (i) networks with linear or hexagonal layout
of cells (see Figures
\ref{fig:fourCellLinear}-\ref{fig:sevenCellHexagonal}), and (ii)
networks with arbitrary layout of cells (see Figure
\ref{fig:sevenCellArbitrary}). We index the cells by positive integers
$1, 2, \ldots, N$, in some arbitrary fashion where $N$ denotes the
number of cells. Two completely dependent co-channel cells are said to
be \textit{neighbours}. Note that, \textit{two completely dependent
  cells operating on different non-overlapping channels are
  \textbf{not} neighbors}. Let $\mathcal{N} = \{1, 2, \ldots, N\}$
denote the set of cells and $\mathcal{N}_i \; (\subset \mathcal{N})$
denote the set of neighboring cells of Cell-$i$ $(i \in
\mathcal{N})$. Note that $i \notin \mathcal{N}_i$. Given the location
of the APs and a channel assignment, we can obtain the set of
neighboring cells for each cell. The key to modeling the cell level
contention is the cell level \textit{contention graph} $\mathcal{G}$
which is obtained by representing every cell by a vertex and joining
every pair of neighbors by an edge. Figures
\ref{fig:fourCellLinear}-\ref{fig:sevenCellArbitrary} also depict the
contention graphs corresponding to each cell topology. 

In Section \ref{subsec:saturated-case}, we model the
case where nodes are infinitely backlogged and are transferring packets
to one or more nodes in the same cell using UDP connection(s). Notice
that single-hop direct communications among the STAs in the same cell
without involving the AP are also allowed. In Section
\ref{subsec:TCP-traffic}, we extend to the case when STAs download
long files through their respective APs using \textit{persistent} TCP
connections. 


\subsection{Modeling with Saturated MAC Queues}
\label{subsec:saturated-case}

Due to the PBD condition, nodes belonging to the same cell have an
identical view of the rest of the network. When one node senses the
medium idle (resp. busy) so do the other nodes in the same cell and we
say that a cell is sensing the medium idle (resp. busy). Since the
nodes are saturated, whenever a cell senses the medium idle, all the
nodes in the cell decrement their back-off 
counters per idle back-off slot that elapses in their \textit{local}
medium\footnote{Nodes belonging to different co-channel cells can have
  different views of the network activity.} and we say that the cell
is in back-off. If the nodes were not saturated, a node with an empty
MAC queue would not count down during the ``medium idle'' periods and
the number of \textit{contending} nodes would be time-varying. With
saturated AP and STA queues, the number of contending nodes in each
cell remains constant. 

We say that a cell transmits when one or more nodes in the cell
transmit(s). When two or more nodes in the same cell transmit, an
\textit{intra-cell} collision occurs. Consider Figure
\ref{fig:fourCellLinear}. There are periods during which all the four
cells are in back-off. We model these periods, when none of the cells
is transmitting, by the state $\Phi$ where $\Phi$ denotes the
\textit{empty set}. The system remains in State $\Phi$ until one or
more cell(s) transmit(s). When a cell transmits, its neighbors sense
the transmission after a propagation delay and they defer medium
access. We then say that the neighbors are \textit{blocked} due to
carrier sensing. However, two neighboring cells can start transmitting
together before they could sense each other's transmissions resulting
in \textit{synchronous inter-cell} collisions. 


We observe that a cell can be in one of the three states: (i)
transmitting, (ii) blocked, or (iii) in back-off. Modeling the
synchronous inter-cell collisions requires a discrete time slotted
model. However, this would require a large state space since the cells
change their states in an asynchronous manner. For example, consider
Figure \ref{fig:fourCellLinear} and suppose that Cell-1 starts
transmitting and blocks Cell-2 after a propagation delay. However,
Cell-3 is independent of Cell-1 and can start transmitting at any
instant during Cell-1's transmission. Thus, the evolution of the
system is partly asynchronous and partly synchronous. To capture both,
we follow a \textit{two-stage} approach along the lines of
\cite{wanet.garetto_etal08starvation}. In the first stage, we ignore
inter-cell collisions and assume that blocking due to carrier sensing
is immediate. We develop a continuous time model as in
\cite{wanet.boorstyn87multihop} to obtain the fraction of
\textit{time} each cell is transmitting/blocked/in back-off. In the
second stage, we obtain the fraction of \textit{slots} in which
various subsets of neighboring cells can start transmitting
together. This would allow us to compute the collision probabilities
accounting for synchronous inter-cell collisions. We combine the above
through a fixed-point equation and compute the throughputs using the
solution of the fixed-point equation. We define the following
\cite{wanet.kumar_etal07new_insights}:

\vspace {2.0mm}

\noindent $n_i :=$ \parbox[t] {8cm} {number of nodes in Cell-$i$}

\vspace {2.0mm}

\noindent $\beta_i :=$ \parbox[t] {7.8cm} {(transmission) attempt
  probability (over the back-off slots) of the nodes in Cell-$i$} 

\vspace {2.0mm}

\noindent $\gamma_i :=$ \parbox[t] {8cm} {collision probability as
  seen by the nodes in Cell-$i$ (conditioned on an attempt being
  made)} 

\vspace {2.0mm}

Let $A_i(T)$ denote the cumulative number of attempts made by a tagged
node (any node due to symmetry) in Cell-$i$ up to time $T$. Let
$C_i(T)$ denote the cumulative number of collisions as seen by the
tagged node up to time $T$. Let $B_i(T)$ denote the cumulative number
of back-off slots elapsed in Cell-$i$ up to time $T$. Then, the
attempt probability $\beta_i$ and the (conditional) collision
probability $\gamma_i$ of the tagged node are more precisely defined
by

\[ \beta_i := \lim_{T \rightarrow \infty} \frac{A_i(T)}{B_i(T)} \; ; \;
\gamma_i := \lim_{T \rightarrow \infty} \frac{C_i(T)}{A_i(T)} \; .\] 

Note that $\beta_i$ is the attempt probability of the nodes in
Cell-$i$, irrespective of whether the nodes in the other cells can
also attempt. This is a simplification and can be viewed as an
extension to the \textit{decoupling approximation} introduced
in \cite{wanet.bianchi00performance}. Using the decoupling
approximation and the analysis in
\cite{wanet.kumar_etal07new_insights}, the attempt probability
$\beta_i$ of the nodes in Cell-$i$, $\forall i \in \mathcal{N}$, can
be related to $\gamma_i$ as

\begin{equation}
  \label{eqn:G_gamma_i}
\beta_i = G(\gamma_i) := \frac{1 + \gamma_i  \ldots + \gamma_i^K}{b_0
  + \gamma_i b_1 \ldots + \gamma_i^k b_k + \ldots + \gamma_i^K b_K} 
\end{equation}

\noindent where $K$ denotes the \textit{retry limit} and $b_k$, $0
\leq k \leq K$, denotes the mean back-off sampled after $k$
collisions.

\textbf{The First Stage:} When Cell-$i$ and some (or all) of its
neighboring cells are in back-off their (cell level) attempt processes
compete until one of the cells, say, Cell-$j$, ${j \in \mathcal{N}_i
  \cup \{i\}}$, transmits. Since, we ignore inter-cell collisions in
the first stage, the possibility of two or more neighboring cells
attempting together is ruled out. When Cell-$i$ wins the
contention, we say that it has become \textit{active}. When Cell-$i$
becomes active, it gains the control over its local medium by
immediately blocking its neighboring cells that are not yet
blocked. We assume that the time until Cell-$i$ goes from the back-off
state to the active state is exponentially distributed with mean
$\frac{1}{\lambda_i}$. The activation rate $\lambda_i$ is given by 


\begin{equation}
\label{eqn:lambda_i-multicell-CTMC}
\lambda_i = \frac{1 - (1 - \beta_i)^{n_i}}{\sigma} 
\end{equation}

\noindent where $\sigma$ denotes the duration of a back-off slot (in
seconds) and $1 - (1 - \beta_i)^{n_i}$ is the probability that there is
an attempt in Cell-$i$ per back-off slot. Notice that we have
converted the aggregate attempt probability in a cell per back-off
\textit{slot} to an attempt rate over back-off \textit{time}. Also
notice that, our assumption of exponential ``time until transition
from the back-off state to the active state'' is the continuous time
analogue of the assumption of geometric ``number of slots until
attempt'' in the discrete time models of
\cite{wanet.cali-etal00throughput-limit} and
\cite{wanet.kumar_etal07new_insights}. 

\begin{discussion}
In \cite{wanet.garetto_etal08starvation}, the authors use an
unconditional activation rate $\lambda$ over all times as well as a
conditional activation rate $g$ over the back-off times and relate the
two rates through a throughput equation which makes their model
unnecessarily complicated. The activation rate $\lambda$ in our model
is conditional on being in the back-off state. Thus, we use a single
activation rate and our model is much simpler than that of
\cite{wanet.garetto_etal08starvation}. Our modified approach can also
be applied to simplify the node level model of
\cite{wanet.garetto_etal08starvation}. \hfill \IEEEQED 
\end{discussion}

When Cell-$i$ becomes active, it remains active and its neighbors
remain blocked due to Cell-$i$ until Cell-$i$'s transmission finishes
and an idle DIFS period elapses. The active periods of Cell-$i$ are of
mean duration $\frac{1}{\mu_i}$. When Cell-$i$ becomes active through
a successful transmission (resp. an intra-cell collision) its
neighbors remain blocked due to Cell-$i$ for a \textit{success time}
$T_s$ (resp. a \textit{collision time} $T_c$)\footnote{For the
  \textit{Basic Access} (resp. \textit{RTS/CTS}) mechanism, $T_s$
  corresponds to the time DATA-SIFS-ACK-DIFS
  (resp. RTS-SIFS-CTS-SIFS-DATA-SIFS-ACK-DIFS) and $T_c$ corresponds
  to the time DATA-DIFS (resp. RTS-DIFS). When DATA payload sizes are
  not fixed, $T_s$ and $T_c$ are to be computed using the expected
  payload size.}. Hence, $\frac{1}{\mu_i}$ is given by 

\begin{eqnarray}
\label{eqn:mu_i-multicell-CTMC}
\frac{1}{\mu_i} &=& \left( \frac{n_i \beta_i (1 - \beta_i)^{n_i -
    1}}{1 - (1 - \beta_i)^{n_i}} \right) \cdot (T_s)
    \nonumber \\ 
&& + \; \left(1 - \frac{n_i \beta_i (1 - \beta_i)^{n_i - 1}}{1 - (1 -
    \beta_i)^{n_i}} \right) \cdot (T_c)
\end{eqnarray}

\noindent where $\displaystyle \frac{n_i \beta_i (1 - \beta_i)^{n_i -
    1}}{1 - (1 - \beta_i)^{n_i}}$ is the probability that Cell-$i$
becomes active through a success given that it becomes active. 

Due to carrier sensing, at any point of time, only a set ${\mathcal{A}
  \; (\subset \mathcal{N})}$ of mutually independent cells can be
active together, i.e., $\mathcal{A}$ must be an \textit{independent
  set} (of vertices) of the contention graph $\mathcal{G}$\footnote{An
  independent set of vertices of a graph $G$ is a set of vertices
  of $G$ such that no two vertices in the set are connected by an
  edge.}. From the contention graph $\mathcal{G}$, we can determine
the set of cells $\mathcal{B}_{\mathcal{A}}$ that get blocked due to
$\mathcal{A}$, and the set of cells $\mathcal{U}_{\mathcal{A}}$ that
remain in back-off, i.e., the set of cells in which nodes can continue
to decrement their back-off counters. Note that $\mathcal{A}$,
$\mathcal{B}_{\mathcal{A}}$ and $\mathcal{U}_{\mathcal{A}}$ form a
partition of $\mathcal{N}$, i.e., $\mathcal{A}$,
$\mathcal{B}_{\mathcal{A}}$ and $\mathcal{U}_{\mathcal{A}}$ are
pairwise disjoint and $\mathcal{N} = \mathcal{A} \cup
\mathcal{B}_{\mathcal{A}} \cup \mathcal{U}_{\mathcal{A}}$.

\begin{figure}[tb]
\centering \
  \begin{minipage}{8cm}
  \begin{center}
\psfig{figure=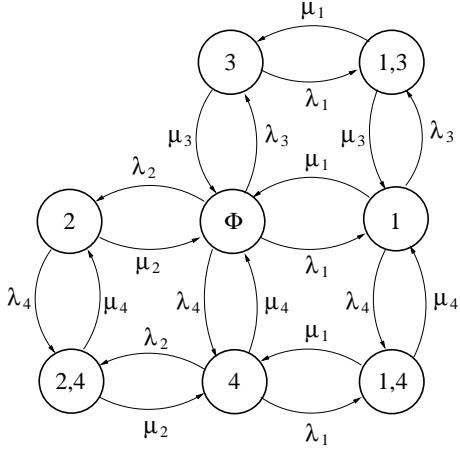,height=6cm,width=6cm}
     \caption{The CTMC describing the cell level contention for the
     four linearly placed cells given in
     Figure \ref{fig:fourCellLinear}\label{fig:CTMCfourcell}.}  
  \end{center}
  \end{minipage}
\end{figure}

We take $\mathcal{A}(t)$, i.e., the set $\mathcal{A}$ of active cells
at time $t$, as the state of the multi-cell system at time $t$. It is
worthwhile now to mention the \textit{insensitivity} result of
Boorstyn et al. \cite{wanet.boorstyn87multihop} which says that the
product-form solution provided by their model is insensitive to the
packet length distribution and depends only on the mean packet
lengths. Applying their insensitivity argument, we take the active
periods of Cell-$i$ to be i.i.d. exponential random variables with mean
$\frac{1}{\mu_i}$. Due to exponential activation rates and active
periods, at any time $t$, the rate of transition to the next state is
completely determined by the current state $\mathcal{A}(t)$. For
example, Cell-$j$, $j \in \mathcal{U}_{\mathcal{A}}$, joins the set
$\mathcal{A}$ (and its neighboring cells that are also in
$\mathcal{U}_{\mathcal{A}}$ join the set $\mathcal{B}_{\mathcal{A}}$)
at a rate $\lambda_j$. Similarly, Cell-$i$, $i \in \mathcal{A}$,
leaves the set $\mathcal{A}$ (and its neighboring cells that are
blocked only due to Cell-$i$ leave the set
$\mathcal{B}_{\mathcal{A}}$) to join the set
$\mathcal{U}_{\mathcal{A}}$ at a rate $\mu_i$. In summary, the process
$\{\mathcal{A}(t), t \geq 0\}$ has the structure of a Continuous Time
Markov Chain (CTMC). This CTMC contains a finite number of states and
is irreducible. Hence, it is stationary and ergodic. The set of all
possible independent sets which constitutes the state space of the
CTMC $\{\mathcal{A}(t), t \geq 0\}$ is denoted by
$\bmath{\mathcal{A}}$. For a given contention graph,
$\bmath{\mathcal{A}}$ can be determined. For the topology given in
Figure \ref{fig:fourCellLinear}, we have $\mathcal{N} = \{1,2,3,4\}$
and $\bmath{\mathcal{A}} =
\{\Phi,\{1\},\{2\},\{3\},\{4\},\{1,3\},\{1,4\},\{2,4\}\}$ where we
recall that $\Phi$ denotes the empty set. The CTMC $\{\mathcal{A}(t),
t \geq 0\}$ corresponding to this example is given in Figure
\ref{fig:CTMCfourcell}. 

It can be checked that the transition structure of the CTMC
$\{\mathcal{A}(t), t \geq 0\}$ satisfies the Kolmogorov Criterion for
reversibility (see \cite{theory.kelly79reversibility}). Hence, the
stationary probability distribution $\pi(\mathcal{A}), \mathcal{A} \in
\bmath{\mathcal{A}}$, satisfies the detailed balance equations,


\[\pi(\mathcal{A})\lambda_i = \pi(\mathcal{A} \cup \{i\})\mu_i \; , \;
(\forall i \in \mathcal{U}_{\mathcal{A}})\]

\noindent and the stationary probability distribution has the form

\begin{equation}
\label{eqn:stationary-probabilities}
\pi(\mathcal{A}) = \left( \prod_{i \in \mathcal{A}}\rho_i \right)
\pi(\Phi), \; \; \; (\forall \mathcal{A} \in \bmath{\mathcal{A}})
\end{equation}

\noindent where $\rho_i := \frac{\lambda_i}{\mu_i}$ and $\pi(\Phi)$
(which denotes the stationary probability that none of the cells is
active) is determined from the normalization equation

\begin{equation}
\label{eqn:normalization-equation}
\sum_{\mathcal{A} \in \bmath{\mathcal{A}}} \pi(\mathcal{A}) = 1.
\end{equation}

Combining Equations \ref{eqn:stationary-probabilities} and
\ref{eqn:normalization-equation}, we obtain

\begin{equation}
\label{eqn:stationary-probabilities-complete}
\pi(\mathcal{A}) = \displaystyle \frac{\left( \displaystyle \prod_{i
    \in \mathcal{A}}\rho_i \right)}{\displaystyle \sum_{\mathcal{A}
    \in \bmath{\mathcal{A}}} \left( \displaystyle \prod_{j \in
    \mathcal{A}}\rho_j \right)}, \; \; \; (\forall \mathcal{A} \in
\bmath{\mathcal{A}}) 
\end{equation}

\noindent \textit{Convention:} A product $\prod$ over an empty index
set is taken to be equal to 1. Notice that, with this convention,
Equation \ref{eqn:stationary-probabilities-complete} holds for
$\pi(\Phi)$ as well.

\begin{figure*}
\begin{equation}
\label{eqn:gamma_i-multicell}
\gamma_i = \frac{\sum_{\mathcal{A} \in \bmath{\mathcal{A}} \; : \; i \in
    \mathcal{U}_{\mathcal{A}}} \pi(\mathcal{A}) \left[1 -
    (1-\beta_i)^{n_i-1} \prod_{j \in \mathcal{N}_i \; : \; j \in
      \mathcal{U}_{\mathcal{A}}} (1-\beta_j)^{n_j}
    \right]}{\sum_{\mathcal{A} \in \bmath{\mathcal{A}} \; : \; i \in
    \mathcal{U}_{\mathcal{A}}} \pi(\mathcal{A})} \; \; \; \; \; \; \;
\; \; \; \; \; (\forall i \in \mathcal{N})
\end{equation}
\rule{180mm}{.3mm}
\end{figure*}

\textbf{The Second Stage:} We now compute the collision probabilities
$\gamma_i$'s accounting for inter-cell collisions. Note that
$\gamma_i$ is conditional on an attempt being made by a node in
Cell-$i$. Hence, to compute $\gamma_i$, we focus only on those states
in which Cell-$i$ can attempt. Clearly, Cell-$i$ can attempt in
State-$\mathcal{A}$ iff it is in back-off in State-$\mathcal{A}$,
i.e., iff $i \in \mathcal{U}_{\mathcal{A}}$. In all such states a node
in Cell-$i$ can incur intra-cell collisions due the other nodes in
Cell-$i$. Furthermore, some (or all) of Cell-$i$'s neighbors might
also be in back-off in State-$\mathcal{A}$. If a neighboring cell,
say, Cell-$j$, $j \in \mathcal{N}_i$, is also in back-off in
State-$\mathcal{A}$, i.e., if $j \in \mathcal{U}_{\mathcal{A}}$, then
a node in Cell-$i$ can incur inter-cell collisions due to the nodes in
Cell-$j$. The collision probability $\gamma_i$ is then given by
Equation \ref{eqn:gamma_i-multicell} (appears at the top of the next
page). A formal derivation of Equation \ref{eqn:gamma_i-multicell} is
provided in Appendix \ref{app:derivation-eqn-gamma_i}.

\textbf{Fixed Point Formulation:} Equations
\ref{eqn:lambda_i-multicell-CTMC}, \ref{eqn:mu_i-multicell-CTMC},
\ref{eqn:stationary-probabilities-complete},
\ref{eqn:gamma_i-multicell} and $\rho_i := \frac{\lambda_i}{\mu_i}$
can express the $\gamma_i$'s as functions of only the
$\beta_i$'s. Together with Equation \ref{eqn:G_gamma_i}, they yield an
$N$-dimensional fixed point equation where we recall that $N$ is the
total number of cells. The $N$-dimensional fixed point equation can be
numerically solved to obtain the collision probabilities $\gamma_i$'s,
the attempt probabilities $\beta_i$'s and the stationary probabilities
$\pi(\mathcal{A})$'s. In all of the cases that we have considered, the
fixed point iterations were observed to converge to the same solutions
irrespective of starting points. However, we have not yet been able to
analytically prove the uniqueness of solutions.

\textbf{Calculating the Throughputs:} The stationary probabilities of
the CTMC $\{\mathcal{A}(t), t \geq0\}$ can provide the fraction of
time $x_i$ for which Cell-$i$ is \textit{unblocked}. A cell is said to
be unblocked when it belongs to either $\mathcal{A}$ or
$\mathcal{U}_{\mathcal{A}}$. Thus, $\forall i \in \mathcal{N}$,

\begin{equation}
\label{eqn:fraction-of-channel-time}
x_i = \sum_{\mathcal{A} \in \bmath{\mathcal{A}}\; : \; i \in
  \mathcal{A} \cup \mathcal{U}_{\mathcal{A}}} \pi(\mathcal{A}).
\end{equation}

\begin{definition}
\label{defn:subgraph-delta}
Let $\mathcal{G}_i$, $i \in \mathcal{N}$, denote the subgraph obtained
by removing Cell-$i$ and its neighboring cells in $\mathcal{N}_i$ from
the contention graph $\mathcal{G}$. For a given contention graph
$\mathcal{G}$, let $\Delta$ be defined as follows:

\begin{equation}
\label{eqn:delta}
\Delta := \sum_{\mathcal{A} \in \bmath{\mathcal{A}}} \left( \prod_{j
  \in \mathcal{A}}\rho_j \right)
\end{equation}

\noindent Let $\Delta_i$ denote the $\Delta$ corresponding to the
subgraph $\mathcal{G}_i$. \hfill \IEEEQED
\end{definition}

An important observation which facilitates the computation of the
$x_i$'s is given by the following theorem.

\begin{theorem}
\label{thm:xi} 

The fraction of time $x_i$ for which Cell-$i$ is unblocked is given by 

\begin{equation}
\label{eqn:xi-in-terms-of-deltai}
x_i = \frac{(1 + \rho_i) \Delta_i}{\Delta}, \; \; \; \; \; \forall i
\in \mathcal{N}. 
\end{equation}
\end{theorem}

\proof{See Appendix \ref{app:derivation-theorem-xi}. \hfill \IEEEQED}

Let $\Theta_i$ denote the aggregate throughput of Cell-$i$ in a given
multi-cell network and let $\Theta_{n_i,singlecell}$ denote the
aggregate throughput of Cell-$i$ if it was an isolated cell containing
$n_i$ nodes. Both $\Theta_i$ and $\Theta_{n_i,singlecell}$ are in
packets/sec. We approximate $\Theta_i$ by  

\begin{eqnarray}
\label{eqn:Theta-multicell}
\Theta_{i} &=& x_i \cdot \Theta_{n_i,singlecell} \nonumber \\
&=& \frac{(1 + \rho_i) \Delta_i}{\Delta} \cdot
\Theta_{n_i,singlecell}, 
\end{eqnarray}

\noindent and $\Theta_i$ divided by $n_i$ gives the per node
throughput $\theta_i$ in Cell-$i$, i.e., $\theta_i =
\frac{\Theta_i}{n_i}$ (packets/sec).

\begin{discussion}
Equation \ref{eqn:Theta-multicell} is justified as follows. If
Cell-$i$ is indeed independent of every other cell in the network, it
is never blocked and does not incur inter-cell collisions. Then, we
have $x_i = 1$ and $\Theta_i = \Theta_{n_i,singlecell}$. However, in
general, Cell-$i$ gets blocked due to its neighbors for a fraction of
time $1 - x_i$ and remains unblocked for a fraction of time $x_i$. If
we ignore the time wasted in inter-cell collisions, the times during
which Cell-$i$ is unblocked would consist only of the back-off slots
and the activities of Cell-$i$ by itself. Thus, we approximate the
aggregate throughput of Cell-$i$, over the times during which it is
unblocked, by $\Theta_{n_i,singlecell}$ and $\Theta_{n_i,singlecell}$
multiplied with $x_i$ gives the aggregate throughput $\Theta_i$ of
Cell-$i$ in the multi-cell network. \hfill \IEEEQED 
\end{discussion}

Clearly, Equation \ref{eqn:Theta-multicell} is an approximation since
the time wasted in inter-cell collisions have been ignored. However,
we prefer to keep the approximation because: (1) it is quite accurate
when compared with the simulations (see Section \ref{sec:results}),
and (2) it can be efficiently computed since $\Theta_{n_i,singlecell}$
follows from a single cell analysis and the $\Delta$ as well as
the $\Delta_i$'s can be computed using efficient algorithms
\cite{wanet.kershenbaum-etal87complex}.

\textbf{Complexity of the Model:} In general, obtaining the state
space $\bmath{\mathcal{A}}$ by searching for all possible independent
sets $\mathcal{A}$ could be computationally expensive and the
complexity grows exponentially with the number of vertices in the
contention graph \cite{wanet.kershenbaum-etal87complex}. For realistic
topologies, where connectivity in the contention graph is related to
distance in the physical network, efficient computation of
$\bmath{\mathcal{A}}$ is possible up to several hundred vertices in
the contention graph \cite{wanet.kershenbaum-etal87complex}. Thus, a
cell level model is extremely helpful in analyzing large-scale WLANs
with hundreds of cells since, unlike a node level model, each vertex
in the contention graph now represents a cell\footnote{Note that, the
  state space $\bmath{\mathcal{A}}$ depends only on the cell topology
  and does not depend on the number of nodes in each cell. It also
  does not depend on whether the nodes use same/different protocol
  parameters.}.

\textbf{Large $\rho$ Regime:} Let $\eta$ (resp. $\eta_i$) denote the
number of Maximum Independent Sets\footnote{A maximum independent set
  of a graph is an independent set of the graph having maximum
  cardinality.} (MISs) of $\mathcal{G}$ (resp. $\mathcal{G}_i$) (see
Definition \ref{defn:subgraph-delta}). Notice that, $\eta_i$ is also
equal to the number of MISs of $\mathcal{G}$ to which Cell-$i$
belongs. From Equation \ref{eqn:stationary-probabilities-complete} it
is easy to see that, as $\rho_i \rightarrow \infty$, $\forall i \in
\mathcal{N}$, we have, 

\[\pi(\mathcal{A}) \rightarrow \left\{ \begin{array}{ll}
  \displaystyle \frac{1}{\eta} & \mbox{if $\mathcal{A}$ is an MIS,}
  \\ 0 & \mbox{otherwise}. \end{array} \right.\]

Also, from Equations \ref{eqn:fraction-of-channel-time},
\ref{eqn:delta} and \ref{eqn:xi-in-terms-of-deltai}, we observe that,
as ${\rho_i \rightarrow \infty}$, $\forall i \in \mathcal{N}$, we have, 

\[x_i \rightarrow \displaystyle \frac{\eta_i}{\eta} \; ,\]

\noindent where we recall that $x_i$ is the fraction of time for which
Cell-$i$ is unblocked. The quantity

\[x_i = \displaystyle \frac{\Theta_i}{\Theta_{n_i,singlecell}}\]

\noindent can be interpreted as the throughput of Cell-$i$, normalized
with respect to Cell-$i$'s single cell throughput. We define the
\textit{normalized network throughput} $\bar{\Theta}$ by

\begin{equation}
\label{eqn:normalized-network-throughput}
\bar{\Theta} := \sum_{i=1}^N x_i. 
\end{equation}

Clearly, as $\rho_i \rightarrow \infty$, $\forall i \in \mathcal{N}$,
the cells that belong to every MIS of the contention graph obtain
normalized throughput 1 and the cells that do not belong to any MIS
obtain normalized throughput 0. Similar observations have also been
made in \cite{wanet.wang-kar05multihop}. We further observe that, as
$\rho_i \rightarrow \infty$ for all $i \in \mathcal{N}$, only an MIS
of cells can be active at any point of time. Since an MIS is always
active, as $\rho_i \rightarrow \infty$, $\forall i \in \mathcal{N}$,
we have,

\[\bar{\Theta} \rightarrow \alpha(\mathcal{G}) \; ,\]

\noindent where $\alpha(\mathcal{G})$ denotes the cardinality of an
MIS of $\mathcal{G}$ and is also called the \textit{independence
  number} of $\mathcal{G}$. Notice that $\alpha(\mathcal{G})$ is a
measure of \textit{spatial reuse} in the network.

\subsection{Extension to TCP Traffic}
\label{subsec:TCP-traffic}

We now extend the analysis of Section \ref{subsec:saturated-case} to
the more realistic case when users access a local proxy server via
\textit{persistent} TCP connections. Our extension to TCP-controlled
long file downloads is based on the single cell TCP-WLAN interaction
model of \cite{wanet.bruno08TCPeqvSatModel}. The model proposed in
\cite{wanet.bruno08TCPeqvSatModel} has been shown to be quite accurate
when: (1) the local proxy server is connected with the AP by a
relatively fast wired LAN such that the AP in the WLAN is the
bottleneck, (2) every STA has a \textit{single} persistent TCP
connection, (3) there are no packet losses due to buffer overflow, (4)
the TCP timeouts are set large enough to avoid timeout expirations due
to Round Trip Time (RTT) fluctuations, and (5) the delayed ACK
mechanism is disabled. We keep the above assumptions in this
paper. 

In \cite{wanet.bruno08TCPeqvSatModel}, the authors propose to model a
single cell having an AP and an arbitrary number of STAs with
long-lived TCP connections by an ``equivalent saturated network''
which consists of a saturated AP and a single saturated STA. ``This
equivalent saturated model greatly simplifies the modeling problem
since the TCP flow control mechanisms are now implicitly hidden and
the total throughput can be computed using the saturation analysis
\cite{wanet.bruno08TCPeqvSatModel}.'' Using the equivalent saturated
model of \cite{wanet.bruno08TCPeqvSatModel}, the analysis of Section
\ref{subsec:saturated-case} can be applied to TCP-controlled long file
downloads, taking $n_i = 2, \forall i \in \mathcal{N}$. We explain the
intuition behind this straightforward extension in the following. 

The equivalent saturated model of \cite{wanet.bruno08TCPeqvSatModel}
is quite accurate in the single cell scenario because of the following
reasons. For TCP-controlled long file transfers in WLANs, the AP has
to serve $n \geq 1$ STA(s) by sending TCP DATA (resp. TCP ACK) packets
to $n_d$ downloading (resp. $n_u$ uploading) STAs ($n_l + n_u =
n$). However, since DCF is \textit{packet fair}, the AP does not get
any prioritized access to the medium. Hence, in steady state, the AP's
MAC queue never becomes empty, i.e., the AP is saturated\footnote{This
  observation holds in steady state even for the $n = 1$ STA case if
  the maximum TCP window is large enough. In fact, for the $n = 1$ STA
  case, in steady state, the STA is also saturated
  \cite{wanet.kumar_etal07new_insights}. Thus, for the $n = 1$ STA
  case, the equivalent saturated model is exact.}. Furthermore, since
it is assumed that there are no buffer losses and no TCP timeouts,
initially, the TCP windows keep growing and, in steady state (i) the
TCP windows stay at their maximum, say, $W_{max}$, for every
connection, and (ii) the sum of the number of packets in the AP and in
the STAs is equal to $nW_{max}$. Most of these packets reside in the
AP queue due to the DCF contention mentioned above and only a few STA
queues remain non-empty at any point of time. It turns out that, the
mean number of STAs with non-empty MAC queues $\approx$ 1.5 for
sufficiently large $n$ \cite{wanet.bruno08TCPeqvSatModel} ($n \geq 5$
suffices)\footnote{In fact, applying the Markov model of
  \cite{wanet.harsha07WiNet}, it can be formally proved that the
  expected number of non-empty STAs is equal to 1.5.}. Thus, the $n$
STAs can be approximated by a single saturated STA and the equivalent
saturated model can still predict the throughputs quite
accurately. In fact, the equivalent saturated model can also
accurately predict the collision probability of the AP which is indeed
saturated\footnote{Our simulations indicate that, the equivalent
  saturated model cannot accurately predict the collision
  probabilities of the STAs.}. 

Our extension to multiple cells is based on our key observations that, 

\begin{enumerate}

\item As in single cells, the AP continues to be saturated in
  multi-cell scenarios as well. 

\item The number of non-empty STAs changes only at the end of
  successful transmissions \cite{wanet.harsha07WiNet}; when the AP
  (resp. a STA) succeeds, this number increases (resp. decreases) by
  one. 

\item Whenever a cell is blocked, its activities are \textit{frozen},
  and hence, the number of non-empty STAs cannot change during the
  blocked periods. Thus, we can strip off the blocked periods and put
  the unblocked periods together over which time the cell would behave
  like a single cell. 

\end{enumerate}

Applying the third observation above, the TCP download throughput of a
cell, say, Cell-$i$, in the multi-cell scenario can be obtained by
multiplying the single cell TCP download throughput of Cell-$i$ with
the fraction of time $x_i$ for which Cell-$i$ is unblocked. Clearly,
due to blocking, the likelihood of TCP timeouts increases. However, in
our simulations, we have observed that, \textit{if the minimum value
  of Retransmission TimeOut (RTO) is set to 200ms (which is the
  default value in ns-2.31), the analytical model is capable of
  predicting AP statistics quite well both in the single cell and the
  multi-cell scenarios}. In fact, cells switch between the blocked and
the unblocked states at a fast time scale of few packet transmission
times which takes only a few milliseconds. Hence, timeout expirations
are rare given that there are no buffer losses\footnote{In certain
  cases, where some cells remain severely blocked, their blocked
  periods can be very long and TCP timeouts can occur for their
  respective connections. As we show in Section \ref{sec:results}, our
  analytical model correctly identifies such severely blocked cells by
  predicting close to zero throughputs for them.}.

\section{Results and Discussion}
\label{sec:results}

We carried out simulations using \textit{ns-2.31} \cite{wanet.ns2}. We
created the example topologies given in Figure
\ref{fig:fourCellLinear}-\ref{fig:sevenCellArbitrary}. We chose the
cell radii and the distances among the cells such that the PBD
condition holds. Nodes were randomly placed within the cells. The
saturated case was simulated with high rate CBR over UDP
connections. For the TCP case, we created one TCP download connection
per STA. Each TCP connection was fed by an FTP source with the TCP
source agent attached directly to the AP to emulate a local proxy
server. The AP buffer was set large enough to avoid buffer losses. The
EIFS deferral and the delayed ACK mechanism were disabled. Each case
was simulated 20 times, each run for 200 sec of ``simulated time''. We
report the results for ``Basic Access''. Similar results were obtained
with ``RTS/CTS''. We took 11 Mbps data rate and packet payloads of
1000 bytes. The function ``\textit{fsolve()}'' of MATLAB was used for
solving the $N$-dimensional fixed point equation.

\begin{table}[t]
\begin{center}
\caption{\label{table:results-single-cell-EIFS-off-alt-TCP} Single
  Cell Results: Columns 2-3 Correspond to $n$ Saturated Nodes. Columns
  4-7 Correspond to TCP Downloads with $n$ STAs} 
\begin{tabular}{||c|c|c|c|c|c|c||}
  \hline
$n$ & $\gamma_{ana}^{sat}$ & $\theta_{ana}^{sat}$ &
  $\gamma_{sim}^{AP}$ & $\gamma_{ana}^{AP}$ & $\theta_{sim}^{AP}$ &
  $\theta_{ana}^{AP}$ \\ 
 & & (pkts/sec) & & & (pkts/sec) & (pkts/sec) \\
  \hline
  \hline
1 & 0 & 801.78 & 0.0578 & 0.0586 & 454.01 & 456.53 \\ 
\hline
2 & 0.0586 & 349.94 & 0.0538 & 0.0586 & 456.06 & 456.53 \\ 
  \hline
3 & 0.1077 & 236.09 & 0.0533 & 0.0586 & 456.09 & 456.53 \\ 
\hline
4 & 0.1473 & 176.63 & 0.0528 & 0.0586 & 456.17 & 456.53 \\ 
  \hline
5 & 0.1812 & 140.29 & 0.0531 & 0.0586 & 456.05 & 456.53 \\ 
\hline
6 & 0.2100 & 115.89 & 0.0530 & 0.0586 & 456.10 & 456.53 \\ 
\hline
7 & 0.2348 & 98.43 & 0.0536 & 0.0586 & 456.88 & 456.53 \\
\hline
8 & 0.2565 & 85.35 & 0.0531 & 0.0586 & 456.97 & 456.53 \\
\hline
10 & 0.2927 & 67.11 & 0.0531 & 0.0586 & 456.02 & 456.53 \\
\hline
\end{tabular} \\
\end{center}
\end{table}

Table \ref{table:results-single-cell-EIFS-off-alt-TCP} summarizes the
results for a single cell. Columns 2-3 (resp. 4-7) correspond to the
saturated case (resp. TCP download case) with $n$ saturated nodes
(resp. 1 AP and $n$ STAs). The analytical results for the saturated
case were obtained using \cite{wanet.kumar_etal07new_insights} and
that for the TCP case were obtained using
\cite{wanet.bruno08TCPeqvSatModel}. These single cell results obtained
from known analytical models serve as the basis of our multi-cell
results. The analytical throughputs per-node (resp. of AP) in Column 3
(resp. Column 7) multiplied with the $x_i$'s obtained from our
multi-cell analysis provide the analytical throughputs per-node
(resp. of AP) in the multi-cell cases (see Equation
\ref{eqn:Theta-multicell}).

\subsection{Results for the Saturated Case}
\label{subsec:results-saturated}

\begin{table}[t]
\begin{center}
\caption{\label{table:results-fourCellLinear-EIFS-off} Results for the
  Four Linearly Placed Cells Given in Figure \ref{fig:fourCellLinear}
  when Each Cell Contains $n=5$ Nodes} 
\begin{tabular}{||c|c|c|c|c|c||} 
\hline
Cell & $\gamma_{sim}$ & $\gamma_{ana}$ & $\theta_{sim}$ &
 $\theta_{ana}$ & $\theta_{\infty}$ \\  
index & & & (pkts/sec) & (pkts/sec) & (pkts/sec) \\
\hline
\hline
1 & 0.2351 & 0.2399 & 94.48 & 97.41 & 93.53 \\
\hline
2 & 0.3005 & 0.3146 & 41.21 & 46.66 & 46.76 \\
\hline
3 & 0.2999 & 0.3146 & 41.66 & 46.66 & 46.76 \\
\hline
4 & 0.2359 & 0.2399 & 93.99 & 97.41 & 93.53 \\
\hline
\end{tabular}
\caption{\label{table:results-fiveCellLinear-EIFS-off} Results for the
  Five Linearly Placed Cells Given in Figure \ref{fig:fiveCellLinear}
  when Each Cell Contains $n=5$ Nodes} 
\begin{tabular}{||c|c|c|c|c|c||} 
\hline
Cell & $\gamma_{sim}$ & $\gamma_{ana}$ & $\theta_{sim}$ &
 $\theta_{ana}$ & $\theta_{\infty}$ \\ 
index & & & (pkts/sec) & (pkts/sec) & (pkts/sec) \\
\hline
\hline
1 & 0.1882 & 0.1897 & 129.35 & 131.35 & 140.29 \\ 
\hline
2 & 0.3321 & 0.3975 & 8.69 & 8.64 & 0 \\ 
\hline
3 & 0.1892 & 0.1925 & 123.35 & 126.41 & 140.29 \\ 
\hline
4 & 0.3321 & 0.3975 & 8.72 & 8.64 & 0 \\ 
\hline
5 & 0.1884 & 0.1897 & 129.31 & 131.35 & 140.29 \\ 
\hline
\end{tabular}
\caption{\label{table:results-sevenCellHexagonal-EIFS-off} Results for
  the Seven Hexagonally Placed Cells Given in Figure
  \ref{fig:sevenCellHexagonal} when Each Cell Contains $n=10$ Nodes} 
\begin{tabular}{||c|c|c|c|c|c||} 
\hline
Cell & $\gamma_{sim}$ & $\gamma_{ana}$ & $\theta_{sim}$ &
 $\theta_{ana}$ & $\theta_{\infty}$ \\ 
index & & & (pkts/sec) & (pkts/sec) & (pkts/sec) \\
\hline
\hline
1 & 0.2335 & 0.8896 & 0.003 & 0.02 & 0 \\ 
\hline
2 & 0.3045 & 0.3158 & 31.97 & 32.35 & 33.56 \\
\hline
3 & 0.3061 & 0.3158 & 31.93 & 32.35 & 33.56 \\
\hline
4 & 0.3055 & 0.3158 & 32.05 & 32.35 & 33.56 \\
\hline
5 & 0.3054 & 0.3158 & 31.86 & 32.35 & 33.56 \\
\hline
6 & 0.3070 & 0.3158 & 32.00 & 32.35 & 33.56 \\
\hline
7 & 0.3058 & 0.3158 & 31.95 & 32.35 & 33.56 \\
\hline
\end{tabular}
\caption{\label{table:results-sevenCellArbitraryUnequalNodes-EIFS-off}
  Results for the Seven Arbitrarily Placed Cells Given in
  Figure \ref{fig:sevenCellArbitrary} when Cell-$i$, $1 \leq i \leq 7$
  Contains $n_i = i+1$ Nodes} 
\begin{tabular}{||c|c|c|c|c|c|c||} 
\hline
Cell & $n_i$ & $\gamma_{sim}$ & $\gamma_{ana}$ & $\theta_{sim}$ &
$\theta_{ana}$ & $\theta_{\infty}$ \\ 
index $i$ & & & & (pkts/sec) & (pkts/sec) & (pkts/sec) \\
\hline
\hline
1 & 2 & 0.0669 & 0.0666 & 320.66 & 325.26 & 349.94 \\
\hline
2 & 3 & 0.1163 & 0.1163 & 216.19 & 219.65 & 236.09 \\
\hline
3 & 4 & 0.2764 & 0.3280 & 12.48 & 12.97 & 0 \\
\hline
4 & 5 & 0.3105 & 0.3318 & 34.29 & 40.20 & 46.76 \\
\hline
5 & 6 & 0.2505 & 0.2585 & 83.77 & 84.92 & 77.26 \\
\hline
6 & 7 & 0.3574 & 0.3787 & 28.67 & 32.40 & 32.81 \\
\hline
7 & 8 & 0.3062 & 0.3139 & 56.87 & 59.21 & 56.90 \\
\hline
\end{tabular}
\end{center}
\vspace*{-2.0em}
\end{table}

\begin{figure}[t]
  \centering
  \begin{minipage}{8cm}
  \begin{center}
    \includegraphics[height=6cm,width=8cm]{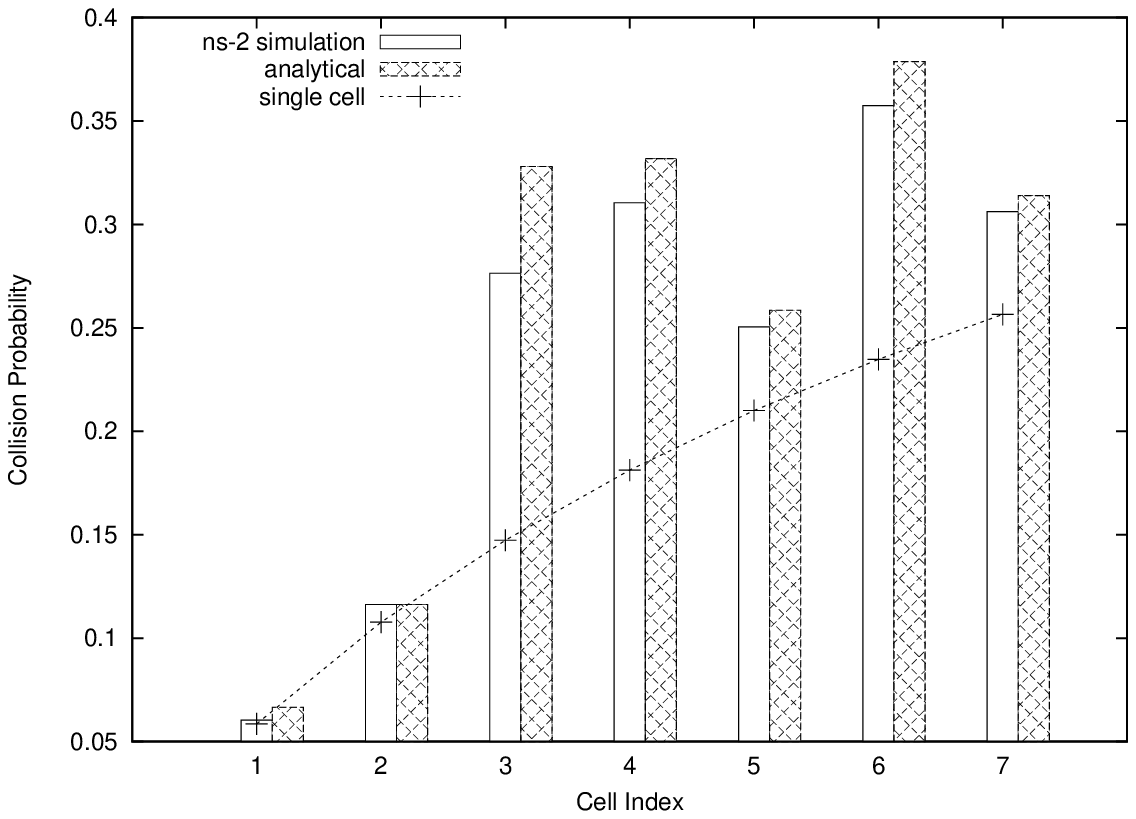}
     \caption{Comparing collision probability $\gamma$ for the example
       scenario in Figure \ref{fig:sevenCellArbitrary} when Cell-$i$,
       $1 \leq i \leq 7$, contains $n_i = i+1$ saturated
       nodes.\label{fig:SevenCellArbitraryGamma}} 
  \end{center}
  \end{minipage}
  \hfill
  \begin{minipage}{8cm}
    \begin{center}
      \includegraphics[height=6cm,width=8cm]{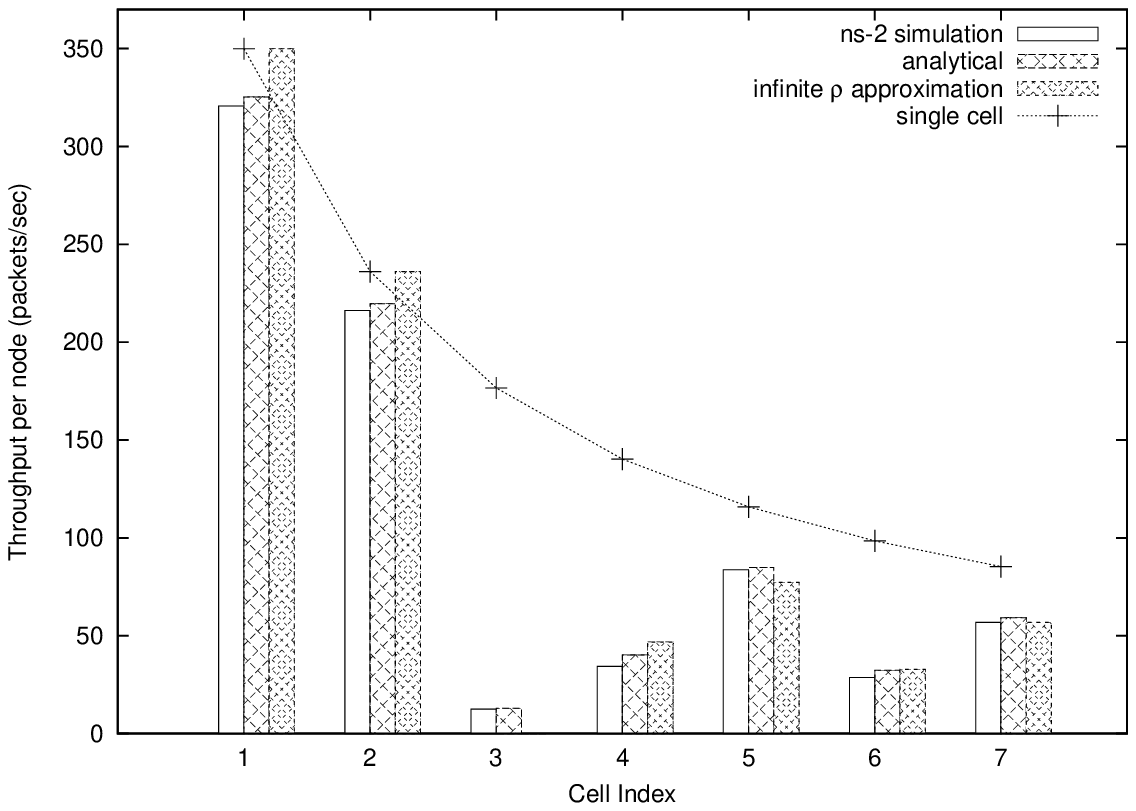}
     \caption{Comparing throughput per node $\theta$ for the example
       scenario in Figure \ref{fig:sevenCellArbitrary} when Cell-$i$,
       $1 \leq i \leq 7$, contains $n_i = i+1$ saturated
       nodes.\label{fig:SevenCellArbitraryTheta}} 
    \end{center}
  \end{minipage}
\end{figure}

Tables
\ref{table:results-fourCellLinear-EIFS-off}-\ref{table:results-sevenCellHexagonal-EIFS-off}
summarize the results for the example multi-cell cases depicted in
Figures \ref{fig:fourCellLinear}-\ref{fig:sevenCellHexagonal},
respectively, when each cell contains $n$ saturated nodes. Table
\ref{table:results-sevenCellArbitraryUnequalNodes-EIFS-off} summarizes
the results for the example case given in Figure
\ref{fig:sevenCellArbitrary} when Cell-$i$, $1 \leq i \leq 7$,
contains $n_i = i+1$ saturated nodes. Quantities denoted with a
subscript \textit{``sim''} (resp. \textit{``ana''}) correspond to
results obtained from \textit{ns-2} simulations (resp. fixed point
analysis). In each case, $\theta_{\infty}$ represents the throughput
per-node obtained by taking $\rho_i \rightarrow \infty, \forall i \in
\mathcal{N}$. We report only the mean values for our simulation
results. The 99\% confidence intervals were observed to be within 5\%
of the mean values. 

We show the plots corresponding
to Table \ref{table:results-sevenCellArbitraryUnequalNodes-EIFS-off}
in Figures \ref{fig:SevenCellArbitraryGamma} and
\ref{fig:SevenCellArbitraryTheta} which compare the collision
probability $\gamma$ and the throughput per node $\theta$,
respectively. In Figures \ref{fig:SevenCellArbitraryGamma} and
\ref{fig:SevenCellArbitraryTheta}, we also show the relevant single
cell results obtained from Table
\ref{table:results-single-cell-EIFS-off-alt-TCP}, i.e., the results
one would expect had the seven cells been mutually
independent. Referring Tables
\ref{table:results-fourCellLinear-EIFS-off}-\ref{table:results-sevenCellArbitraryUnequalNodes-EIFS-off},
and Figures \ref{fig:SevenCellArbitraryGamma} and
\ref{fig:SevenCellArbitraryTheta}, we make the following observations:

\noindent \textbf{O-1.)} Collision probabilities (resp. throughputs)
in the multi-cell scenarios are always higher (resp. lower) than the
corresponding single cell values (see Figures
\ref{fig:SevenCellArbitraryGamma} and
\ref{fig:SevenCellArbitraryTheta}) because (a) inter-cell collisions
can be significant, and (b) due to inter-cell blocking, cells get
opportunity to transmit only a fraction of time. 

\noindent \textbf{O-2.)} Our analytical model is quite accurate (less
than 10\% error in most cases) in predicting the collision
probabilities and throughputs. However, our model always
over-estimates the throughputs since the time wasted in inter-cell
collisions have been ignored. Ignoring inter-cell collisions in the
first stage of the model also over-estimates the fraction of time
spent in back-off. Thus, the collision probabilities are also
over-estimated. 

\noindent \textbf{O-3.)} The relative mismatch between the analytical
model and the simulation is observed to be the worst for cells that
remain blocked most of the time. For example, consider the first row
of Table \ref{table:results-sevenCellHexagonal-EIFS-off} which
corresponds to Cell-1 in Figure \ref{fig:sevenCellHexagonal}. Since
Cell-1 is dependent with respect to every other cell, it obtains very
few attempt opportunities in the simulations and its collision
probability had to be averaged over very few samples. The same
argument applies to the collision probability for Cell-3 in Figure
\ref{fig:SevenCellArbitraryGamma}. 


\noindent \textbf{O-4.)} Our analytical model correctly identifies the
severely blocked cells, e.g., Cell-2 and Cell-4 in Table
\ref{table:results-fiveCellLinear-EIFS-off}, Cell-1 in Table
\ref{table:results-sevenCellHexagonal-EIFS-off}, and Cell-3 in Table
\ref{table:results-sevenCellArbitraryUnequalNodes-EIFS-off} obtain
significantly less throughputs than the other cells in the respective
networks. Furthermore, our model works well with either equal or
unequal number of nodes per cell. 

\noindent \textbf{O-5.)} Throughput distribution among the cells can
be very unfair even over long periods of time. Furthermore,
introduction of a new co-channel cell can drastically alter the
throughput distributions. For example, compare Tables
\ref{table:results-fourCellLinear-EIFS-off} and
\ref{table:results-fiveCellLinear-EIFS-off}. Cell-2 and Cell-4
severely get blocked if Cell-5 is introduced to the four cell network
given in Figure \ref{fig:fourCellLinear}. 

\noindent \textbf{O-6.)} The throughput of a cell cannot be accurately
determined based only on the \textit{number} of interfering
cells. Consider, for example, Figure
\ref{fig:sevenCellArbitrary}. Cell-3 and Cell-4 each have two
neighbors but their per node throughputs $\theta$ are quite different
(see Figure \ref{fig:SevenCellArbitraryTheta}). In particular,
$\theta_4 > \theta_3$ even though $n_3 = 4 < n_4 = 5$. This is due to
Cell-7 which blocks Cell-6 for certain fraction of time during which
Cell-4 gets opportunity to transmit whereas Cell-1 and Cell-2 are
almost never blocked and Cell-3 is almost always blocked due to Cell-1
and Cell-2. Thus, \textit{topology plays the key role and heuristic
  methods based only on the number of neighbors as in
  \cite{wanet.villegas_etal08WiComMobCom} would fail to capture the
  individual cell throughputs}.

\subsection{Results for the TCP Download Case}
\label{subsec:results-TCP-download}

Referring Columns 4-7 of Table
\ref{table:results-single-cell-EIFS-off-alt-TCP} it can be seen that
The AP statistics is largely insensitive to the number of STAs. Also, 

\noindent \textbf{O-7.)} \textit{The collision probability of the AP
  in a single cell with any number of STAs is approximately equal to
  the collision probability in a single cell with two saturated
  nodes}. This can be verified by comparing Columns 4 and 5 with Row 2
Column 2 in Table \ref{table:results-single-cell-EIFS-off-alt-TCP}. 

\noindent \textbf{O-8.)} The AP throughput does not change with the
number of STAs. In fact, the AP throughput with any number of
downloading STAs is equal to the per node throughput in a single cell
containing two saturated nodes with payload size
$\frac{L_{TCP-DATA}+L_{TCP-ACK}}{2}$ where $L_{TCP-DATA}$ and
$L_{TCP-ACK}$ denote the size of TCP DATA and TCP ACK packets. 

\begin{figure}[t]
  \centering
  \begin{minipage}{8cm}
  \begin{center}
    \includegraphics[height=6cm,width=8cm]{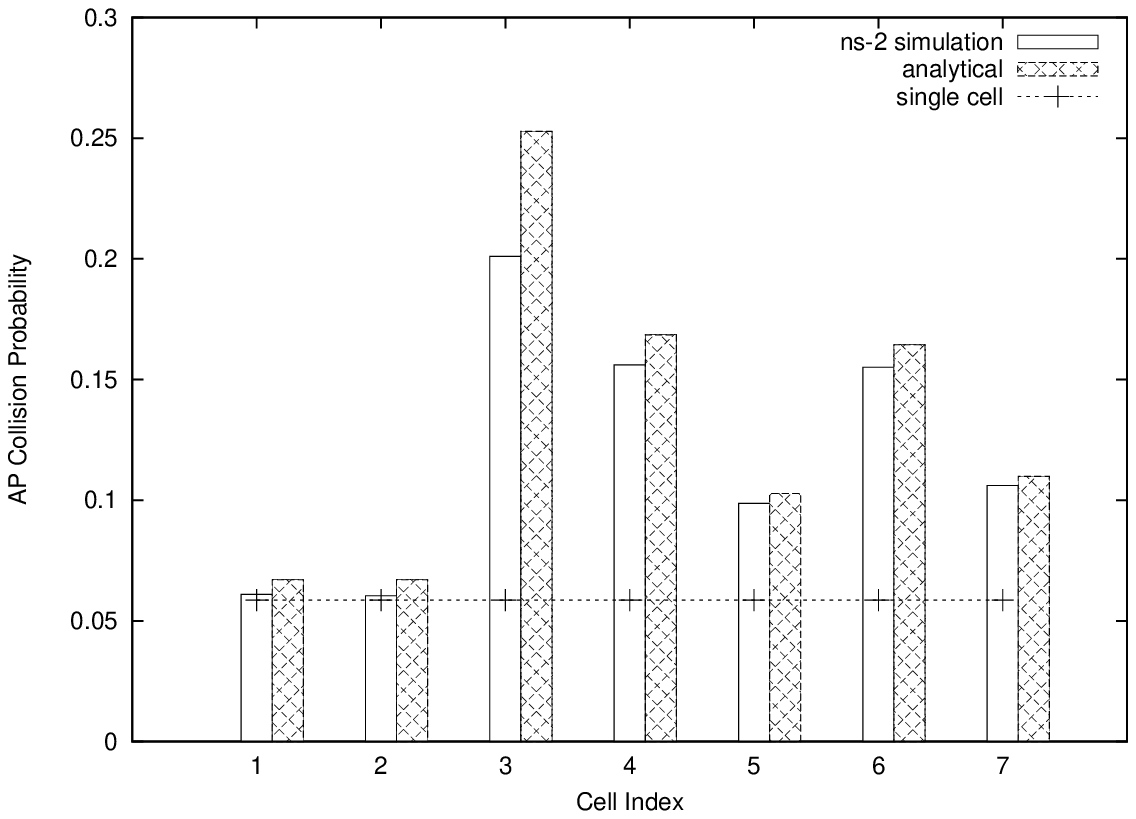}
     \caption{Comparing collision probability $\gamma$ for the example
       scenario in Figure \ref{fig:sevenCellArbitrary} when each cell
       contains an AP and $n = 5$ STAs. STAs are downloading long
       files through their respective APs using TCP
       connections.\label{fig:SevenCellArbitraryGammaTCP}} 
  \end{center}
  \end{minipage}
  \hfill
  \begin{minipage}{8cm}
    \begin{center}
      \includegraphics[height=6cm,width=8cm]{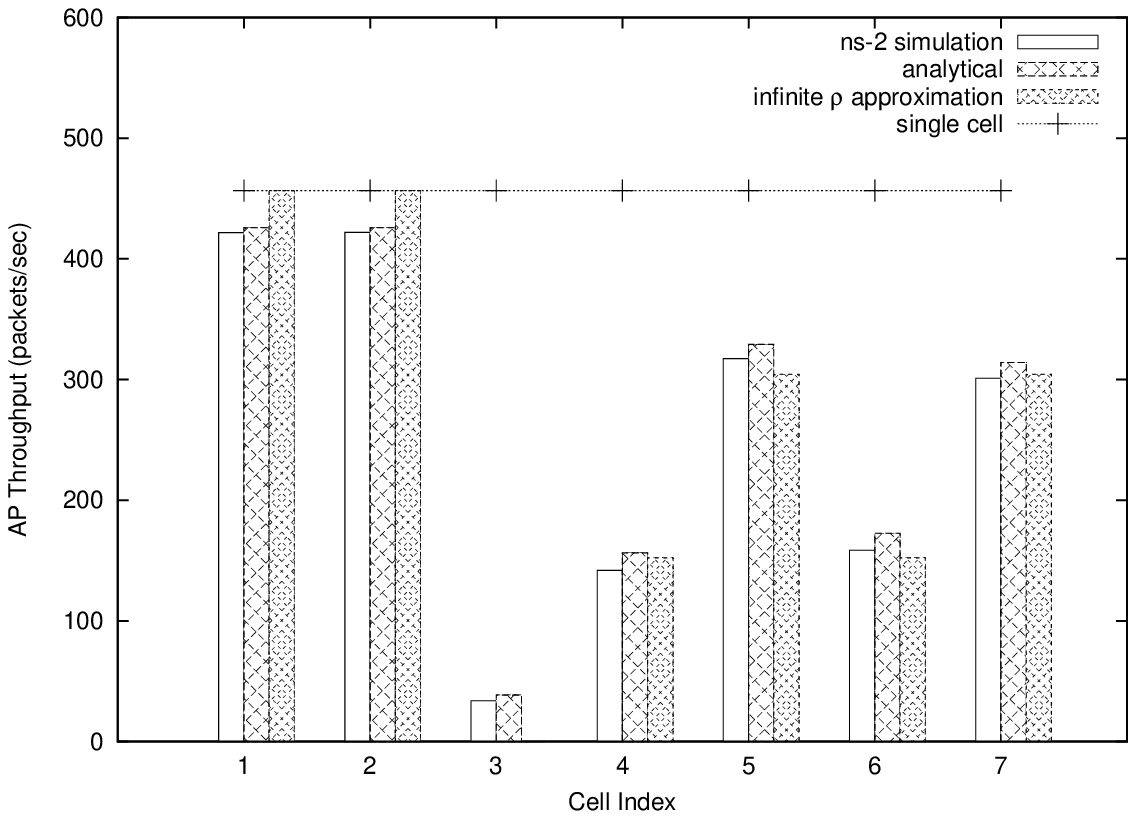}
     \caption{Comparing throughput per node $\theta$ for the example
       scenario in Figure \ref{fig:sevenCellArbitrary} when each cell
       contains an AP and $n = 5$ STAs. STAs are downloading long
       files through their respective APs using TCP
       connections.\label{fig:SevenCellArbitraryThetaTCP}} 
    \end{center}
  \end{minipage}
\end{figure}

Observations \textbf{O-7} and \textbf{O-8} are well-known and they
form the basis for the equivalent saturated model of
\cite{wanet.bruno08TCPeqvSatModel}. Observation \textbf{O-7} has led
to the conclusion in \cite{wanet.ergin08exptIntercellInterference}
that the collision probability of the nodes in a WLAN containing $m$
mutually interfering (i.e., dependent) APs is equal to the collision
probability in a single cell containing $2m$ saturated nodes since
each cell can be assumed to consist of two saturated nodes (a
saturated AP and a saturated STA). Tables
\ref{table:results-fourCellLinear-EIFS-off-alt-TCP}-\ref{table:results-sevenCellArbitrary-EIFS-off-alt-TCP}
summarize the AP statistics for the topologies in Figures
\ref{fig:fourCellLinear}, \ref{fig:fiveCellLinear} and
\ref{fig:sevenCellArbitrary}, respectively. Since the AP statistics
does not change with the number of STAs, we report the results with $n
= 5$ STAs in each case. We show the plots corresponding
to Table \ref{table:results-sevenCellArbitrary-EIFS-off-alt-TCP} in
Figures \ref{fig:SevenCellArbitraryGammaTCP} and
\ref{fig:SevenCellArbitraryThetaTCP} which compare the collision
probability $\gamma$ and the throughput per node $\theta$,
respectively. Referring Tables
\ref{table:results-fourCellLinear-EIFS-off-alt-TCP}-\ref{table:results-sevenCellArbitrary-EIFS-off-alt-TCP},
and Figures \ref{fig:SevenCellArbitraryGammaTCP} and
\ref{fig:SevenCellArbitraryThetaTCP}, we conclude that the foregoing
observations (\textbf{O.1-O.6}) for the saturated case carry over to
TCP-controlled long file transfers as well. Furthermore, we generalize
the conclusion of \cite{wanet.ergin08exptIntercellInterference} as
follows:

\begin{table}[t]
\begin{center}
\caption{\label{table:results-fourCellLinear-EIFS-off-alt-TCP} Results
  for the AP Corresponding to Figure \ref{fig:fourCellLinear} when
  Each Cell Contains 1 AP and $n=5$ STAs} 
\begin{tabular}{||c|c|c|c|c|c||} 
\hline
Cell & $\gamma_{sim,AP}$ & $\gamma_{ana,AP}$ & $\theta_{sim,AP}$ &
 $\theta_{ana,AP}$ & $\theta_{\infty,AP}$ \\  
index & & & (pkts/sec) & (pkts/sec) & (pkts/sec) \\
\hline
\hline
1 & 0.1038 & 0.1033 & 306.33 & 318.73 & 304.35 \\
\hline
2 & 0.1560 & 0.1574 & 153.16 & 169.18 & 152.18 \\
\hline
3 & 0.1555 & 0.1574 & 153.06 & 169.18 & 152.18 \\
\hline
4 & 0.1038 & 0.1033 & 306.41 & 318.73 & 304.35 \\
\hline
\end{tabular}
\caption{\label{table:results-fiveCellLinear-EIFS-off-alt-TCP} Results
  for the AP Corresponding to Figure \ref{fig:fiveCellLinear} when
  Each Cell Contains 1 AP and $n=5$ STAs} 
\begin{tabular}{||c|c|c|c|c|c||} 
\hline
Cell & $\gamma_{sim,AP}$ & $\gamma_{ana,AP}$ & $\theta_{sim,AP}$ &
 $\theta_{ana,AP}$ & $\theta_{\infty,AP}$ \\ 
index & & & (pkts/sec) & (pkts/sec) & (pkts/sec) \\
\hline
\hline
1 & 0.0728 & 0.0775 & 381.21 & 387.16 & 456.53 \\ 
\hline
2 & 0.1793 & 0.1950 & 75.16 & 85.62 & 0 \\ 
\hline
3 & 0.0744 & 0.0832 & 340.24 & 346.47 & 456.53 \\ 
\hline
4 & 0.1786 & 0.1950 & 75.23 & 85.62 & 0 \\ 
\hline
5 & 0.0728 & 0.0775 & 381.15 & 387.16 & 456.53 \\ 
\hline
\end{tabular}
\caption{\label{table:results-sevenCellArbitrary-EIFS-off-alt-TCP}
  Results for the AP Corresponding to
  Figure \ref{fig:sevenCellArbitrary} when Each Cell Contains 1 AP and
  $n=5$ STAs} 
\begin{tabular}{||c|c|c|c|c|c||} 
\hline
Cell & $\gamma_{sim,AP}$ & $\gamma_{ana,AP}$ & $\theta_{sim,AP}$ &
 $\theta_{ana,AP}$ & $\theta_{\infty,AP}$ \\ 
index & & & (pkts/sec) & (pkts/sec) & (pkts/sec) \\
\hline
\hline
1 & 0.0610 & 0.0670 & 421.70 & 425.83 & 456.53 \\
\hline
2 & 0.0604 & 0.0670 & 421.92 & 425.83 & 456.53 \\
\hline
3 & 0.2010 & 0.2528 & 33.79 & 38.50 & 0 \\ 
\hline
4 & 0.1561 & 0.1685 & 141.80 & 156.41 & 152.18 \\ 
\hline
5 & 0.0987 & 0.1028 & 317.39 & 329.06 & 304.35 \\ 
\hline
6 & 0.1551 & 0.1644 & 158.55 & 172.64 & 152.18 \\ 
\hline
7 & 0.1061 & 0.1099 & 301.15 & 314.10 & 304.35 \\ 
\hline
\end{tabular}
\end{center}
\end{table}

\noindent \textbf{O-9.)} The collision probability in a WLAN depends,
not only on the number of interfering APs but also on the fraction of
time for which the neighboring APs can cause collisions. In general,
collision probability does not grow as twice the number of interfering
APs and depends on the cell topology.

Referring Tables
\ref{table:results-fourCellLinear-EIFS-off-alt-TCP}-\ref{table:results-sevenCellArbitrary-EIFS-off-alt-TCP}
we extend the validity of the equivalent saturated model of
\cite{wanet.bruno08TCPeqvSatModel} as follows:

\noindent \textbf{O-10.)} The equivalent saturated model of
\cite{wanet.bruno08TCPeqvSatModel} proposed in the context of a single
cell, preserves its desirable properties, i.e., it predicts the AP
statistics quite well when extended to a multi-cell WLAN that
satisfies the PBD condition.

\subsection{Variation with $\rho$}
\label{subsec:variation-with-rho}

\begin{figure}[t]
  \centering \
  \begin{minipage}{8cm}
    \begin{center}
      \psfig{figure=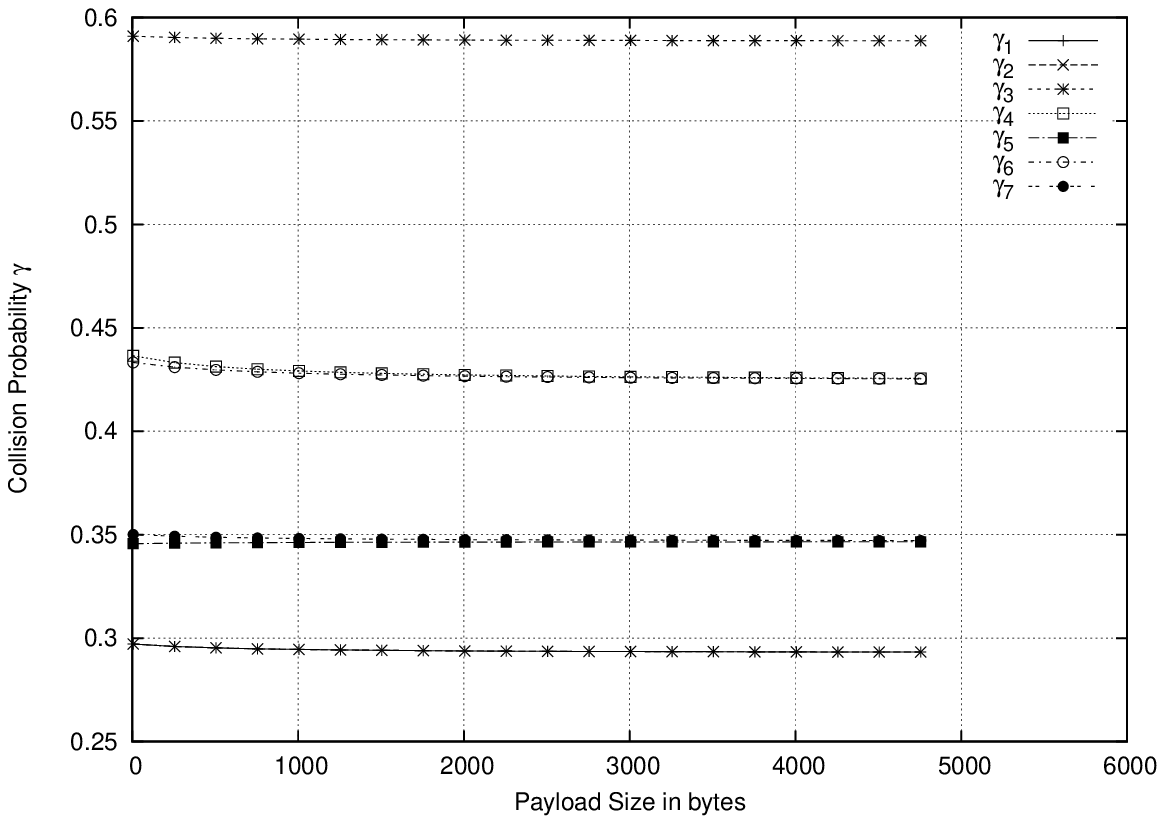,height=6cm,width=8cm}
      \caption{Variation of collision probabilities with payload size
        for the seven cell network in Figure
        \ref{fig:sevenCellArbitrary} when each cell contains $n = 10$
        saturated nodes. \label{fig:GammavariationWithRhoIndividual}} 
    \end{center}
  \end{minipage}
  \begin{minipage}{8cm}
    \begin{center}
      \psfig{figure=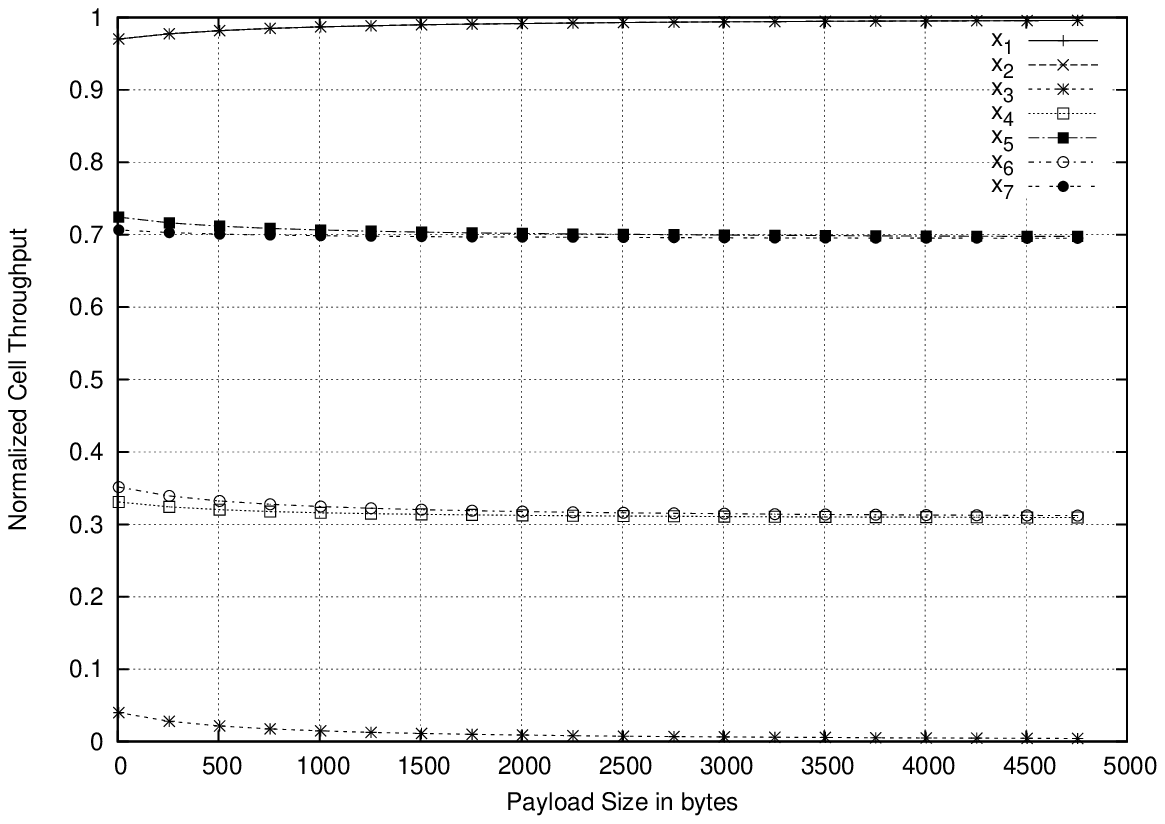,height=6cm,width=8cm}
      \caption{Variation of normalized cell throughputs with payload
        size for the seven cell network in Figure
        \ref{fig:sevenCellArbitrary} when each cell contains $n = 10$
        saturated nodes. \label{fig:variationWithRhoIndividual}} 
    \end{center}
  \end{minipage}
\end{figure}

Till now, we have discussed the results corresponding to the payload
size of 1000 bytes. We now examine how the results vary
with the payload size. Figures
\ref{fig:GammavariationWithRhoIndividual} and
\ref{fig:variationWithRhoIndividual} depict the variation with payload
size of analytically computed collision probabilities and normalized
cell throughputs, respectively, for the seven cell network of Figure
\ref{fig:sevenCellArbitrary} when each cell contains $n = 10$
saturated nodes. Similar results were obtained for the TCP case as
well. From Figures \ref{fig:GammavariationWithRhoIndividual} and
\ref{fig:variationWithRhoIndividual}, we observe that: 

\noindent \textbf{O-11.)} The results are largely insensitive to the
variation in payload size. Moreover, as the payload size increases,
the normalized cell throughputs become closer to the normalized
throughputs under the infinite $\rho$ approximation, i.e., they become
closer to $x_1 = x_2 = 1, x_3 = 0, x_4 = x_6 = \frac{1}{3}, x_5 = x_7
= \frac{2}{3}$. Hence, except for very small payload sizes, the
infinite $\rho$ approximation can be expected to provide fairly
accurate predictions. Furthermore, for sufficiently large payload
sizes, we have, $\bar{\Theta} \approx \alpha(\mathcal{G})$.

\section{A Simple Design Example}
\label{sec:design-example}

\begin{figure}[t]
\centering \
  \begin{minipage}{8cm}
  \begin{center}
\psfig{figure=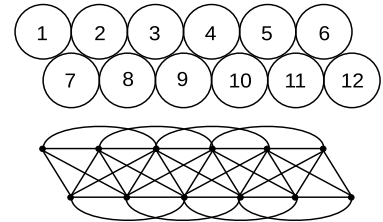,height=4cm,width=5.5cm}
     \caption{A 12-cell network: Cell indices have been marked. Also
       shown is the cell level \textit{physical} contention graph
       assumed for the network. \label{fig:12cellNetwork}} 
  \end{center}
  \end{minipage}
\end{figure}

\begin{figure*}[tbh]
\centering \
  \begin{minipage}{5cm}
  \begin{center}
\includegraphics[height=3cm,width=5cm]{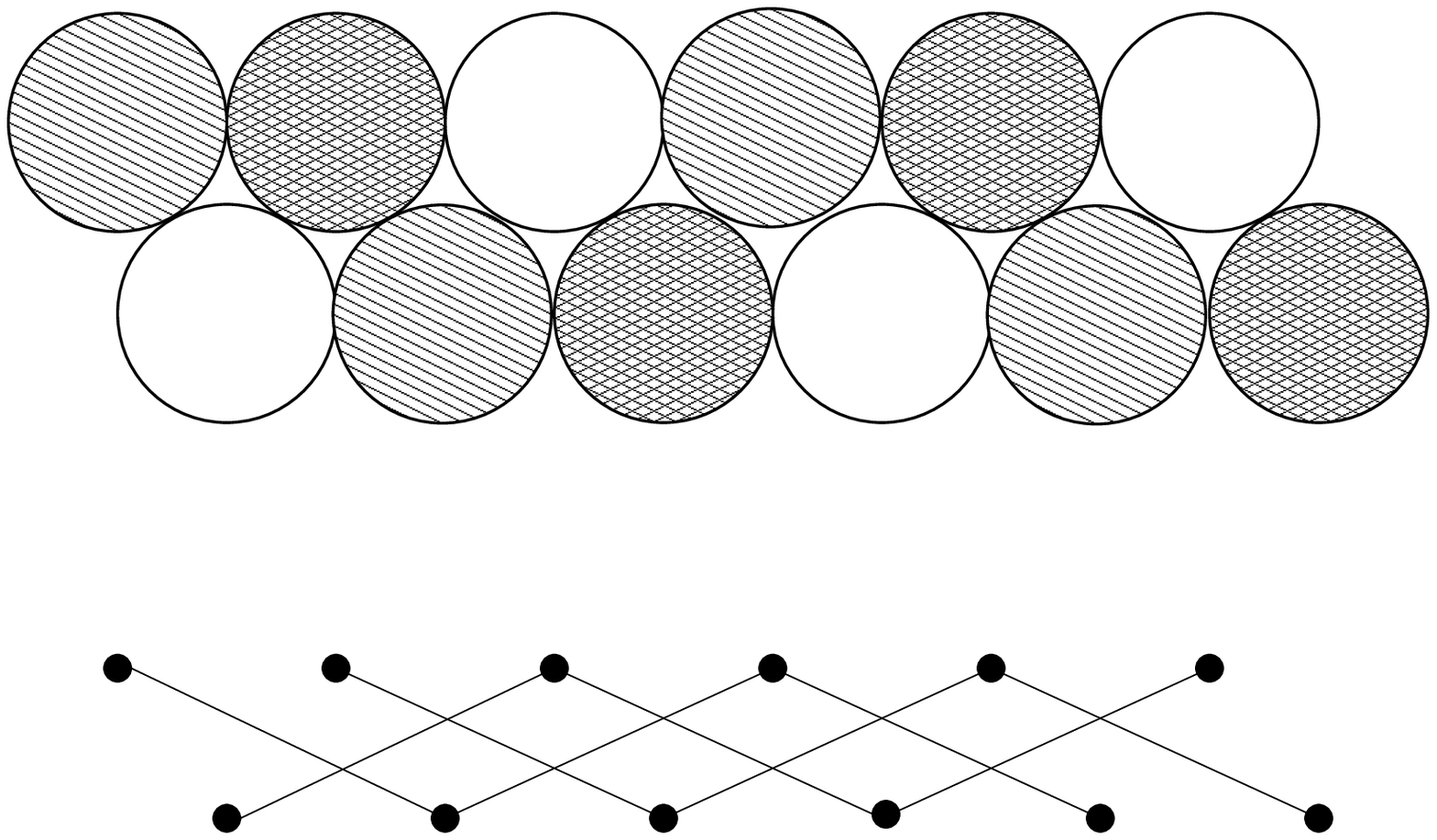}
\subfigure[]{\label{fig:allocationOne}}
  \end{center}
  \end{minipage}
\hfill
  \begin{minipage}{5cm}
  \begin{center}
\includegraphics[height=3cm,width=5cm]{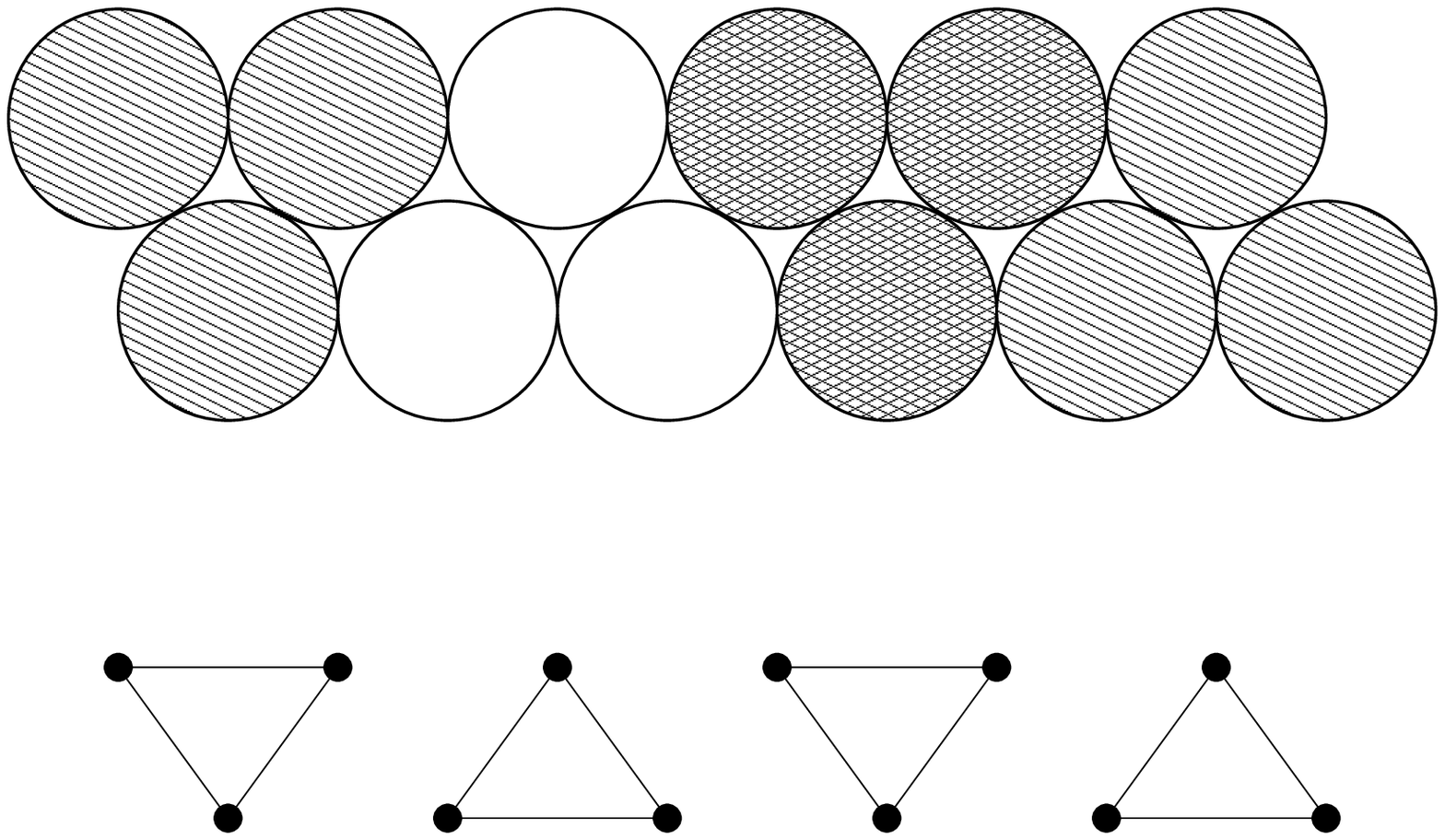}
\subfigure[]{\label{fig:allocationTwo}}
  \end{center}
  \end{minipage}
\hfill
  \begin{minipage}{5cm}
  \begin{center}
\includegraphics[height=3cm,width=5cm]{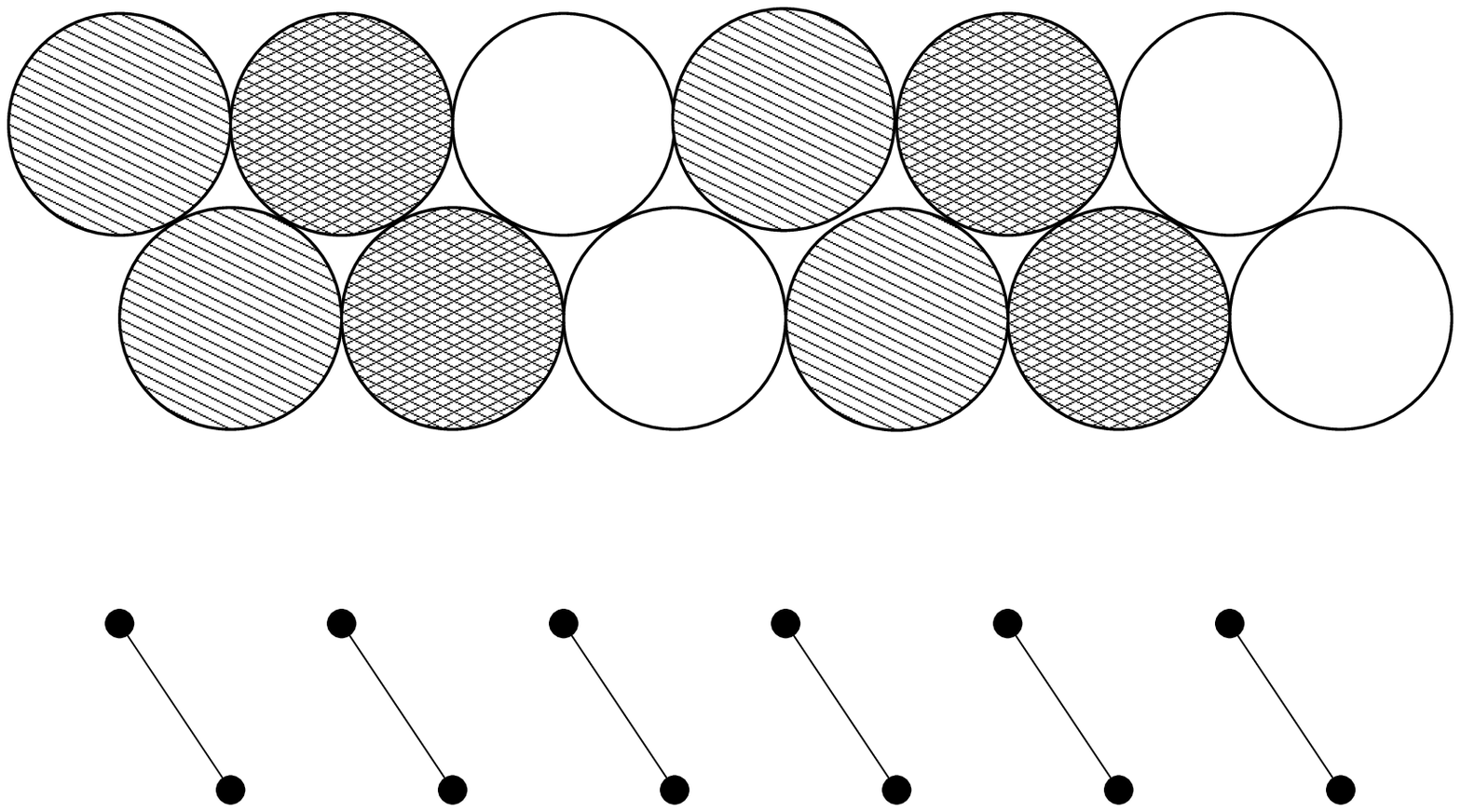}
\subfigure[]{\label{fig:allocationThree}}
  \end{center}
 \end{minipage}
\caption{Comparing three channel assignments for a 12-cell network:
  The cell level \textit{logical} contention graph are also shown for
  each assignment.} 
\end{figure*}

We consider a 12-cell network as shown in Figure
\ref{fig:12cellNetwork} and apply our throughput model to compare
three different channel assignments as shown in Figures
\ref{fig:allocationOne}-\ref{fig:allocationThree}. Notice that the
``logical'' contention graphs for the three assignments in Figures
\ref{fig:allocationOne}-\ref{fig:allocationThree} are different from
the ``physical'' contention graph of the network shown in Figure
\ref{fig:12cellNetwork} which is based on the physical separation
among the cells. Two physically dependent cells become logically
independent if they are assigned different channels and the
corresponding edge will be missing in the logical contention
graph. Our objective is to determine the best, among the three
assignments, in terms of the normalized network throughput
$\bar{\Theta}$ (see Equation \ref{eqn:normalized-network-throughput})
and the fairness index reported in
\cite{cs.jain_etal84fairness}. Considering fairness at the cell level,
the fairness index $J$ is given by \cite{cs.jain_etal84fairness} 

\vspace{-1mm}

\[J := \displaystyle \frac{(\sum_{i=1}^N x_i)^2}{N (\sum_{i=1}^N
  x_i^2)} \; ,\]

\vspace{-1mm}

\noindent where we recall that $x_i$ is the normalized throughput of
Cell-$i$. Note that, $0 \leq J \leq 1$, and higher values of $J$ imply
more fairness. 

Observe that co-channel cells in the first assignment have topology as
in Figure \ref{fig:fourCellLinear}. There are three subsystem of
co-channel cells each consisting of four cells. The infinite $\rho$
approximation predicts that, the two cells at the middle (resp. at the
end) for each subsystem of co-channel cells would obtain normalized
throughput $\approx \frac{1}{3}$ (resp. $\approx \frac{2}{3}$). The
normalized network throughput for the first assignment is $6$ and the
fairness index is $0.45$. In the second assignment, each cell obtains
normalized throughput $\approx \frac{1}{3}$. The normalized network
throughput for the second assignment is $4$ and the fairness index is
$1$. Thus, the fairness has attained its maximum in the second
assignment. However, this assignment is not desirable since
$\bar{\Theta}$ for this assignment is only $\frac{2}{3}$ of
$\bar{\Theta}$ in the first assignment. In the third assignment, each
cell obtains normalized throughput $\approx \frac{1}{2}$. The
normalized network throughput for the third assignment is $6$ and the
fairness index is equal to $1$. Clearly, the third assignment is the
best one among the three. 

Note that, an SINR based model is likely to prefer the first
assignment to the third. However, interference by nodes within
$R_{cs}$ can be taken care by carrier sensing, and hence, such
interference should not be given undue weights. If required, cells
that are very close to each other may be assigned the same channel to
form ``big'' cells. Unlike a cell which covers a large area by a
single AP and provides connectivity at low PHY rates to STAs that are
located at the edge of the cell, a cluster of mutually interfering
small co-channel cells each containing an AP to ensure connectivity at
high PHY rates can provide high capacity and fairness. This emphasizes
the impact of carrier sensing in 802.11 networks. 


\section{Applying the Analytical Model along with a Learning
  Algorithm} 
\label{sec:channel-assignment-algorithm}

In the 12-cell network of Figure \ref{fig:12cellNetwork}, with 3
non-overlapping channels, we need to examine $3^{12}$ possibilities to
determine the optimal assignment. It is also not clear if the
normalized network throughput for the 12-cell network can be actually
more than 6 with some other assignment. Under the infinite $\rho$
approximation, the maximum value of $\bar{\Theta}$ is equal to the
maximum independence number over all possible logical contention
graphs. Clearly, the problem of finding an assignment that maximizes
$\bar{\Theta}$ is NP-hard and, in general, it is not desirable to
examine all the possibilities. 

We now demonstrate how our analytical model could be applied along
with a Learning Automata (LA) algorithm called the Linear
Reward-Inaction $(L_{R-I})$ algorithm \cite{FAP.sastryetal94LRIgames}
to obtain optimal channel assignments. For every cell $i \in
\mathcal{N}$, we maintain an $M$-dimensional probability vector
$\bmath{p}_i$ where $M$ denotes the number of available channels. The
$(L_{R-I})$ algorithm proceeds in steps. Let $k = 0, 1, 2, \ldots$ be
the step indices. At each step $k$, a channel is selected for
Cell-$i$, $i \in \mathcal{N}$, according to the probability
distribution


\[\bmath{p}_i(k) := \Big(p_{i,1}(k), p_{i,2}(k), \ldots,
p_{i,M}(k)\Big),\]


\noindent where  


\[p_{i,j}(k) := \mbox{ \parbox[t] {6.7cm} {the probability that
    Channel-$j$ is selected for Cell-$i$ at step $k$.}}\] 


Let $\bmath{c}(k) = \Big(c_1(k), c_2(k), \ldots, c_N(k)\Big)$ be the
channel assignment at step $k$, where 


\[c_i(k) := \mbox{the channel selected for Cell-$i$ at step $k$}. \]


Given a channel assignment $\bmath{c}$, we can obtain the logical
contention graph $\mathcal{G}(\bmath{c})$ and applying our throughput
model, we can compute the normalized throughput $x_i(\bmath{c})$ of
every cell $i$ with the assignment $\bmath{c}$. We define the
\textit{sum utility} for an assignment $\bmath{c}$ by 


\begin{equation}
\label{eqn:utility-assignment}
U(\bmath{c}) := \sum_{i=1}^N u(x_i(\bmath{c})),
\end{equation}


\noindent where $u(\cdot)$ is some suitably defined \textit{increasing
  concave} function. For a given $u(\cdot)$, we compute the sum
utility at step $k$ by 


\[U(\bmath{c}(k)) = \sum_{i=1}^N u(x_i(\bmath{c}(k))) \; ,\]


\noindent and apply the $L_{R-I}$ algorithm which consists of the
following steps \cite{FAP.sastryetal94LRIgames}:

\begin{enumerate}

\item Begin with $p_{i,j} \in (0,1)$, $\forall i \in \mathcal{N}$, $j
  = 1, 2, \ldots, M$, such that $\bmath{p}_i \cdot \bmath{1} = 1$,
  where $\bmath{1}$ is the $M$-dimensional vector with all components
  equal to 1. 

\item Update the probability vectors of Cell-$i$, $\forall i \in
  \mathcal{N}$, as 


\[\bmath{p}_i(k+1) = \bmath{p}_i(k) + b \; U(\bmath{c}(k))
\Big(\bmath{\delta}_{c_i(k)} - \bmath{p}_i(k)\Big) \; ,\]


\noindent where $\bmath{\delta}_j$ denotes the $M$-dimensional
probability vector with unit mass on Channel-$j$ and $0 < b < 1$. 

\end{enumerate}

The parameter $b$ is called the \textit{learning parameter} or the 
\textit{step-size parameter} which determines the convergence
properties of the algorithm; with smaller $b$ convergence is slower
but the algorithm may not converge to the desired optimum if $b$ is
not small enough \cite{FAP.sastryetal94LRIgames}. Notice that, to
ensure non-negativity of the probability vectors after every update
with any $b \in (0,1)$, the sum utility $U$ must satisfy $0 \leq U
\leq 1$. To maximize $\bar{\Theta} = \sum_{i=1}^N x_i$, we take the
\textit{average} normalized network throughput

\begin{equation}
\label{eqn:average-normalized-network-throughput}
U_{\bar{\Theta}} := \displaystyle \frac{1}{N} \sum_{i=1}^N x_i
\end{equation}

\noindent as the sum utility $U$ where we compute the $x_i$'s applying
the infinite $\rho$ approximation. Note that, $0 \leq U_{\bar{\Theta}}
\leq 1$. Other utility functions can also be considered provided that
$0 \leq U \leq 1$. Define

\[\mathbf{P} := (\bmath{p}_1, \bmath{p}_2, \ldots, \bmath{p}_N). \]

Using the terminology of \cite{FAP.sastryetal94LRIgames}, the super
vector $\mathbf{P}$ will be called a \textit{strategy} for the channel
assignment problem. A strategy $\mathbf{P}$ such that, $\forall i \in
\mathcal{N}$, $p_{i,j} = 1$ for some $j$, $j = 1, 2, \ldots, M$, will
be called a \textit{pure strategy}. A strategy $\mathbf{P}$ that is
not a pure strategy is called a \textit{mixed strategy}. Under the
$L_{R-I}$ algorithm, $\{\mathbf{P}(k), k \geq 0\}$ is a Markov process
with the pure strategies as the only absorbing states. Thus, in
practice, the $L_{R-I}$ algorithm always converges to a pure strategy
rather than to a mixed strategy
\cite{FAP.sastryetal94LRIgames}. Furthermore, by Theorem 4.1 of
\cite{FAP.sastryetal94LRIgames}, the $L_{R-I}$ algorithm always
converges to one of the \textit{Nash equilibria}. Thus, in practice,
channel assignment by the $L_{R-I}$ algorithm always provides an
assignment $\bmath{c}^*$ which is one of the \textit{Nash equilibria
  in pure strategies} in the sense that 

\[U(\bmath{c}^*) \geq U(\bmath{c}),\]

\noindent for all $\bmath{c} $ that differs from $\bmath{c}^*$ by
exactly one element, i.e., changing the channel of one of the cells in
the assignment $\bmath{c}^*$ does not increase the sum utility $U$.

\subsection{Results for Channel Assignment by the $L_{R-I}$ Algorithm} 
\label{subsec:results-LRI}

\begin{figure}[tb]
\centering \
  \begin{minipage}{8cm}
  \begin{center}
\psfig{figure=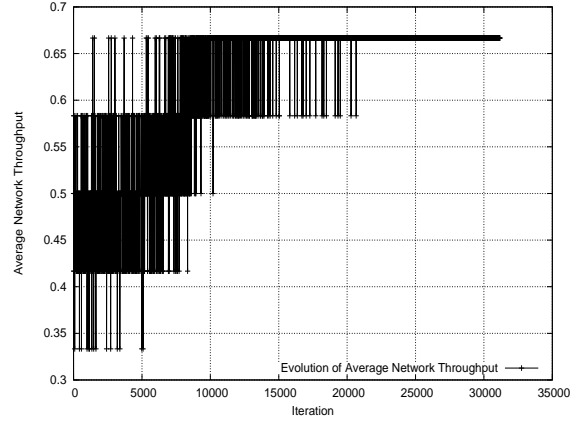,height=6cm,width=8cm}
     \caption{Evolution of the average normalized network
       throughput $U_{\bar{\Theta}}$ under the $L_{R-I}$ algorithm for
       the 12-cell network in Figure \ref{fig:12cellNetwork} with $M =
       3$ channels, $b = 0.01$, and $p_{ij} = \frac{1}{3}$, $\forall i
       \in \mathcal{N}$ and $j = 1, 2, \ldots,
       M$. \label{fig:NwTheta12CellSumTheta}} 
\end{center}
  \end{minipage}
\end{figure}



\begin{figure}[t]
\centering \
  \begin{minipage}{8cm}
  \begin{center}
\psfig{figure=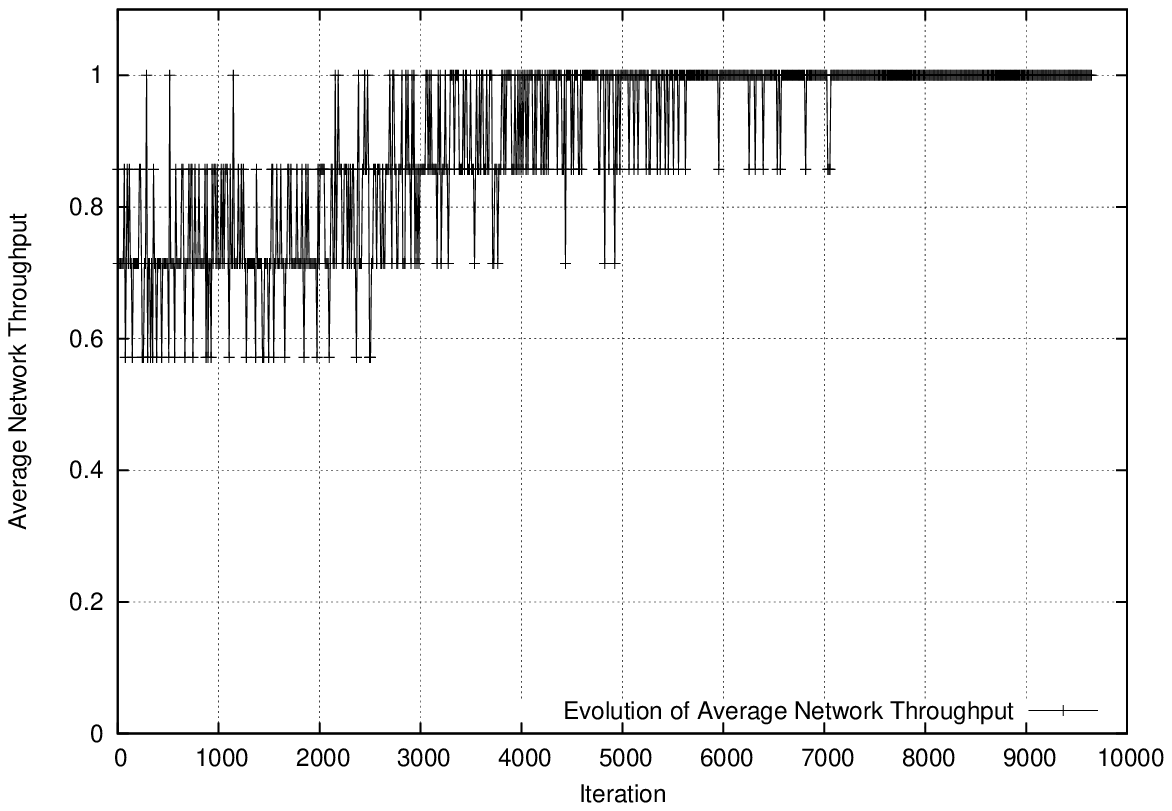,height=6cm,width=8cm}
     \caption{Evolution of the average normalized network throughput
       $U_{\bar{\Theta}}$ for the 7-cell example in Figure
       \ref{fig:sevenCellArbitrary} with $M=2$ channels, $b = 0.01$,
       and $p_{i1} = p_{i2} = 0.5$, $i = 1, 2, \ldots, 7$. In this
       example, the $L_{R-I}$ algorithm converges to a global
       optimum. \label{fig:NwTheta7CellAlpha01Begin5050EndGlobalLRI}} 
  \end{center}
  \end{minipage}
  \begin{minipage}{8cm}
  \begin{center}
\psfig{figure=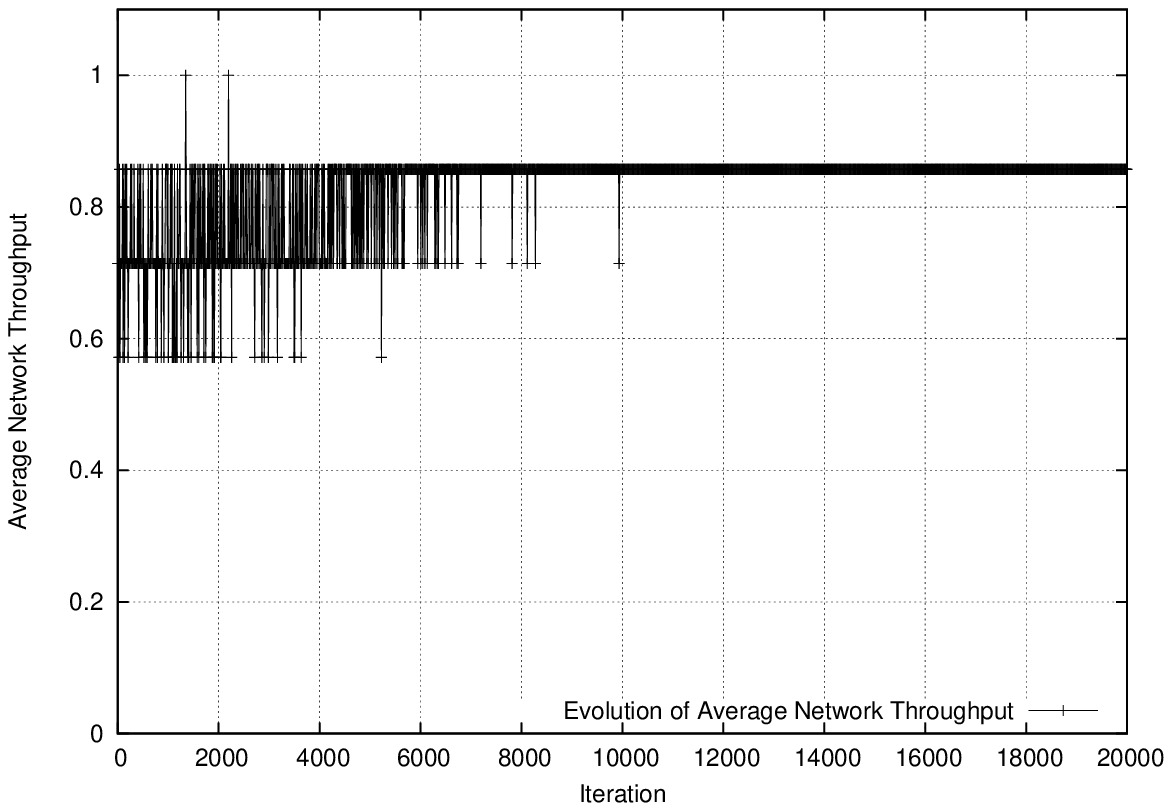,height=6cm,width=8cm}
     \caption{Evolution of the average normalized network throughput
       $U_{\bar{\Theta}}$ for the 7-cell example in Figure
       \ref{fig:sevenCellArbitrary} with $M=2$ channels, $b = 0.01$,
       and $p_{i1} = p_{i2} = 0.5$, $i = 1, 2, \ldots, 7$. In this
       example, the $L_{R-I}$ algorithm converges to a local
       optimum.\label{fig:NwTheta7CellAlpha01Begin5050EndLocalConvergedLRI}} 
  \end{center}
  \end{minipage}
\end{figure}

\begin{figure}[t]
\centering \
  \begin{minipage}{8cm}
  \begin{center}
\psfig{figure=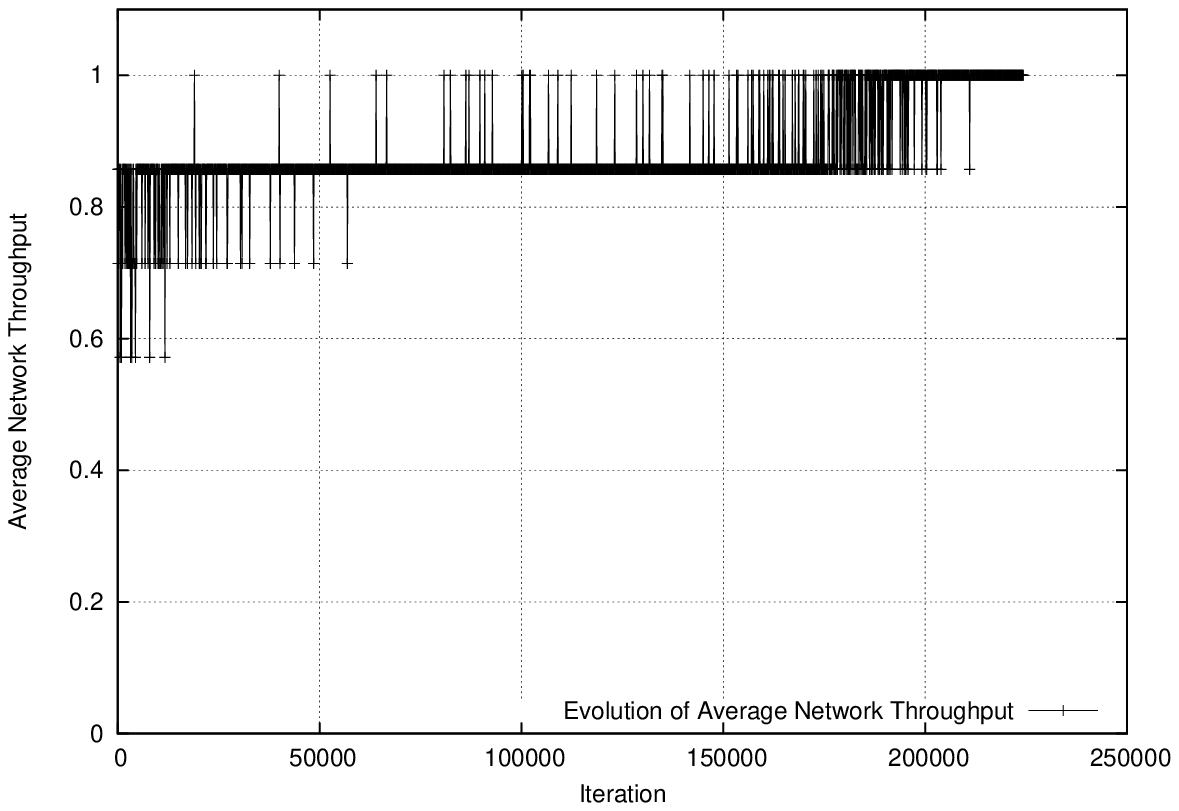,height=6cm,width=8cm}
     \caption{Evolution of the average normalized network throughput
       $U_{\bar{\Theta}}$ for the 7-cell example in Figure
       \ref{fig:sevenCellArbitrary} with $M=2$ channels, $b = 0.001$,
       and $p_{11} = p_{21} = p_{32} = p_{42} = p_{51} = p_{61} =
       p_{72} = 0.8$. In this example, the $L_{R-I}$ algorithm
       converges to a global
       optimum.\label{fig:NwTheta7CellAlpha001Begin8020EndGlobalLRI}} 
  \end{center}
  \end{minipage}
  \begin{minipage}{8cm}
  \begin{center}
\psfig{figure=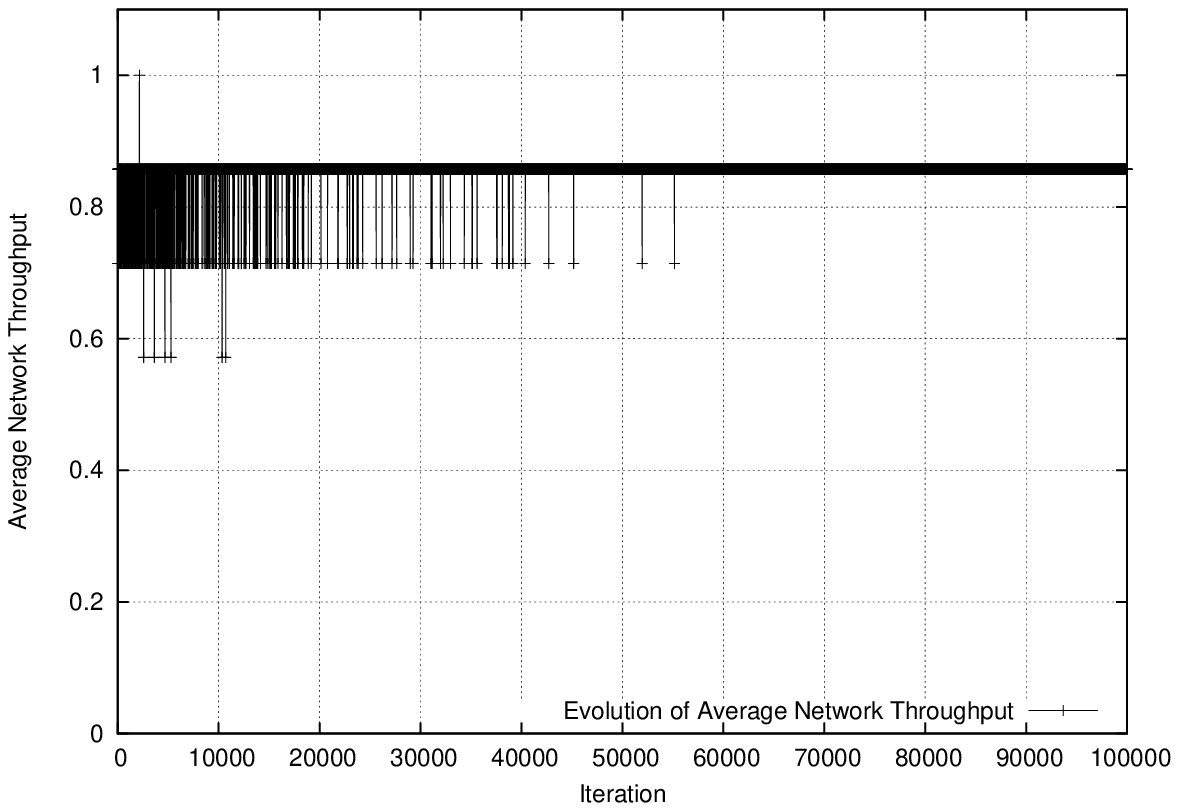,height=6cm,width=8cm}
     \caption{Evolution of the average normalized network throughput
       $U_{\bar{\Theta}}$ for the 7-cell example in Figure
       \ref{fig:sevenCellArbitrary} with $M=2$ channels, $b = 0.001$,
       and $p_{11} = p_{21} = p_{32} = p_{42} = p_{51} = p_{61} =
       p_{72} = 0.9$. In this example, the $L_{R-I}$ algorithm
       converges to a local
       optimum.\label{fig:NwTheta7CellAlpha001Begin9010EndLocalLRI}} 
  \end{center}
  \end{minipage}
\end{figure}

Figure \ref{fig:NwTheta12CellSumTheta} shows the evolution of the
average normalized network throughput $U_{\bar{\Theta}}$ under the
$L_{R-I}$ algorithm with $M = 3$ channels and $b = 0.01$ for the
12-cell network in Figure \ref{fig:12cellNetwork}. We began with
uniform \textit{unbiased} probability vectors $\bmath{p}_i =
(\frac{1}{3}, \frac{1}{3}, \frac{1}{3})$, $\forall i \in
\mathcal{N}$. The algorithm converged to the assignment\footnote{We do
  not provide the plots showing evolution of the $p_{i,j}$'s to save
  space.} $c_1 = c_6 = c_9 = 1, {c_4 = c_7 = c_{12} = 3}, c_2 = c_3 =
c_5 = c_8 = c_{10} = c_{11} = 2$. Notice that $U_{\bar{\Theta}}$
converges to $\frac{2}{3}$ which corresponds to $\bar{\Theta} =
8$. It can be verified that this is the maximum possible value that
$\bar{\Theta}$ can take for the given network with 3
channels\footnote{Note that permutations of channels do not change the
  logical topology and the associated $\bar{\Theta}$'s would be
  equal. For example, $c_1 = c_6 = c_9 = 2, c_4 = c_7 = c_{12} = 1,
  c_2 = c_3 = c_5 = c_8 = c_{10} = c_{11} = 3$ would also give
  $\bar{\Theta} = 8$.}. We found that the $L_{R-I}$ algorithm always converges to
a globally optimum solution if started with uniform unbiased
probability vectors and if $b$ is sufficiently small. However,
starting with probability vectors that are highly \textit{biased}
towards some assignment or if $b$ is not small enough, it may not
converge to a globally optimum solution. We demonstrate this through
Figures
\ref{fig:NwTheta7CellAlpha01Begin5050EndGlobalLRI}-\ref{fig:NwTheta7CellAlpha001Begin9010EndLocalLRI}
for the 7-cell example in Figure \ref{fig:sevenCellArbitrary} with $M
= 2$ channels.


Beginning with unbiased probability vectors and $b = 0.01$ the
algorithm may converge to a global optimum with $U_{\bar{\Theta}} = 1$
or $\bar{\Theta} = 7$ (Figure
\ref{fig:NwTheta7CellAlpha01Begin5050EndGlobalLRI}). There are two
possible global optima, namely, $c_1 = c_2 = c_4 = c_7 = 1, c_3 = c_5
= c_6 = 2$ and $c_1 = c_2 = c_4 = c_7 = 2, c_3 = c_5 = c_6 = 1$. In
Figure \ref{fig:NwTheta7CellAlpha01Begin5050EndLocalConvergedLRI}, we
show that, with $b = 0.01$, the algorithm may also converge to a
solution that is not a global optimum. In fact, for the specific
simulation run, the algorithm converged to the Nash equilibrium $c_1 =
c_2 = c_5 = c_6 = 1, c_3 = c_4 = c_7 = 2$ with $\bar{\Theta} = 6$
which is not a global optimum. The assignment $c_1 = c_2 = c_5 = c_6 =
1, c_3 = c_4 = c_7 = 2$ is a Nash equilibrium in the sense that
changing the channel of one of the cells does not increase
$\bar{\Theta}$, and hence, the sum utility $U_{\bar{\Theta}}$. With $b
= 0.001$, and beginning with unbiased uniform probability vectors, we
observed that the algorithm always converged to a global optimum. With
$b = 0.001$, the algorithm converged to a global optimum even when we
began with probability vectors biased towards the assignment $c_1 =
c_2 = c_5 = c_6 = 1, c_3 = c_4 = c_7 = 2$ (Figure
\ref{fig:NwTheta7CellAlpha001Begin8020EndGlobalLRI}). However, when
the initial probability vectors were highly biased towards the
assignment $c_1 = c_2 = c_5 = c_6 = 1, c_3 = c_4 = c_7 = 2$, the
algorithm converged to the biased assignment which is not a global
optimum (Figure \ref{fig:NwTheta7CellAlpha001Begin9010EndLocalLRI}).

\section{A Simple and Fast Algorithm for Maximizing Normalized Network
  Throughput}
\label{sec:simple-algorithm}

We observe that the $L_{R-I}$ algorithm takes a large number of
iterations to converge and guarantees convergence only to Nash
equilibria. Clearly, for the 7-cell example with $M=2$ channels, an
\textit{exhaustive search} would have required examination of only
$2^7 = 128$ possibilities whereas the $L_{R-I}$ algorithm takes many
more steps to converge. The Linear Reward-Penalty ($L_{R-P}$)
algorithm of \cite{wanet.leith_clifford06CFL} and the
\textit{simulated annealing} algorithm of
\cite{wanet.kauffmann07selfOrganization} guarantee convergence to a
globally optimum solution as the number of iterations goes to
infinity. A \textit{greedy} version of simulated annealing algorithm
in \cite{wanet.kauffmann07selfOrganization} is relatively faster but
still takes a large number of iterations to converge and guarantees
convergence only to locally optimum solutions. To maximize the
normalized network throughput $\bar{\Theta}$ we now propose a simple
and fast decentralized algorithm which can be easily implemented in
real networks. 

We form a physical contention graph $\mathcal{G}$ in which every
completely dependent pair of cells are neighbors (since we have not
yet assigned channels) and our objective is to transform $\mathcal{G}$
to a logical contention graph $\mathcal{G}(\bmath{c})$ by an
assignment $\bmath{c}$ so that $\alpha(\mathcal{G}(\bmath{c}))$ is
maximized. As noted in Observation \textbf{O.11}, for sufficiently
large packet sizes, we have $\bar{\Theta} \approx
\alpha(\mathcal{G})$. Hence, maximizing
$\alpha(\mathcal{G}(\bmath{c}))$ would maximize $\bar{\Theta}$. \\

\textbf{Maximal Independent Set Algorithm (mISA) :} We propose the
following channel assignment algorithm: (recall that $M$ is the number
of available channels)

(1) Choose a \textit{maximal} independent set (mIS)\footnote{The lower
  case `m' corresponds to ``maximal'' as opposed to the upper case `M'
  which corresponds to ``maximum''.} of cells, assign them Channel-1
and remove them from the graph. 

(2) Increment the channel index and repeat the procedure on the
\textit{residual} graph until one channel is left. 

(3) Assign Channel-$M$ to all the cells in the residual graph after
$M-1$ steps. 

Notice that mISA takes only $M$ steps. Also notice that, the residual
graph may become \textit{null} (i.e., it might not have any vertices
left) after $M' < M$ steps. Clearly, mISA is based on a classical
graph coloring technique, but the novelty lies in our recognition of
the notion of optimality that mISA provides which is given by the
following

\begin{theorem} The channel assignments by mISA are Nash equilibria in
  pure strategies for the objective of maximizing normalized network
  throughput $\bar{\Theta}$ as $\rho_i \rightarrow \infty$, $\forall i
  \in \mathcal{N}$. Furthermore, mISA provides a globally optimum
  solution if $M \geq D(\mathcal{G}) + 1$ where $D(\mathcal{G})$
  denotes the maximum vertex degree in the physical contention graph
  $\mathcal{G}$. 
\end{theorem}

\proof{If the residual graph becomes null in less than $M$ steps, then
  every cell would have a channel different from all its neighbors'
  and, we have, $\bar{\Theta} = N$, i.e., mISA would provide a
  globally optimum solution. Hence, to prove the first part of the
  theorem, we assume, without loss of generality, that the residual
  graph after $M-1$ steps is not null. 

  Suppose that $N_j$ cells are assigned Channel-$j$ in Step-$j$,
  $1 \leq j \leq M-1$. Let $\mathcal{G}_{j}$ be the residual graph
  after $j$ steps, $1 \leq j \leq M - 1$. Let $\bar{\Theta}_k$
  denote the aggregate normalized throughput of the cells on
  Channel-$k$, $1 \leq k \leq M$. Then, we have, $\bar{\Theta}_j =
  N_j$, $1 \leq j \leq M-1$, due to independence, and $\bar{\Theta}_M
  = \alpha(\mathcal{G}_{M-1})$. Hence, $\bar{\Theta} = \sum_{k=1}^M
  \bar{\Theta}_k = \sum_{j=1}^{M-1} N_j +
  \alpha(\mathcal{G}_{M-1})$. Suppose that a cell on Channel-$j$, $j
  \neq M$, is moved to Channel-$k$, $k \neq j$. Then, $\bar{\Theta}_j$
  decreases by 1 but $\bar{\Theta}_k$ can increase by at most
  1. Hence, $\bar{\Theta}$ cannot increase. Suppose now that a cell on
  Channel-$M$ is moved to Channel-$j$, ${1 \leq j \leq M-1}$. Clearly,
  any cell on Channel-$M$ is dependent (in the original graph
  $\mathcal{G}$) w.r.t. at least one of the $N_j$ cells on Channel-$j$
  since the $N_j$ cells that are already on Channel-$j$, $1 \leq j
  \leq M-1$, form an mIS. Hence, $\bar{\Theta}_j$ does not change but
  $\bar{\Theta}_M$ can only decrease. Hence, $\bar{\Theta}$ cannot
  increase by changing the channel of \textit{only} one of the cells
  and the first part of the theorem is proved. 

  For the second part of the theorem, notice that, every cell in the
  residual graph $\mathcal{G}_{j}$ after $j$ steps, $1 \leq j \leq M -
  1$, must be dependent (in the original graph $\mathcal{G}$)
  w.r.t. at least one cell on each channel $l$, $1 \leq l \leq j -
  1$. This follows from the maximality of the independent sets of
  cells chosen in each of the first $M - 1$ steps. In particular,
  every cell in the residual graph $\mathcal{G}_{M-1}$ is dependent
  (in the original graph $\mathcal{G}$) w.r.t. at least one cell on
  each channel $j$, $1 \leq j \leq M-1$. Since different channels are
  assigned in each step, the vertex degree of every cell in
  $\mathcal{G}_{M-1}$ restricted to the cells in $\mathcal{G}_{M-1}$
  alone would be ${\leq D(\mathcal{G}) - (M-1)}$. If $M \geq
  D(\mathcal{G}) + 1$, then the cells in $\mathcal{G}_{M-1}$ would
  form an independent set and every cell would have a channel
  different from its neighbors'. Thus, if ${M \geq D(\mathcal{G}) +
    1}$, then mISA would provide an assignment with $\bar{\Theta} =
  N$, i.e., a globally optimum solution. \hfill \IEEEQED}

\textbf{Implementation of mISA:} mISA can be implemented in a
decentralized manner as follows. APs sample random back-offs using a
contention window $W$ and contend for accessing the medium using
Channel-1. When an AP wins the contention it keeps transmitting
broadcast packets separated by Short Inter Frame Space (SIFS) for some
duration $T >> \sigma W$ where we recall that $\sigma$ is the duration
of a back-off slot. This emulates the infinite $\rho$ situation since
and AP after wining the contention does not relinquishes the control
over its local medium. We had observed that, as $\rho_i \rightarrow
\infty$, $\forall i \in \mathcal{N}$, only the cells that belong to an
MIS obtain non-zero normalized throughputs. But this holds only in an
\textit{ensemble average} sense. If an AP, after wining the
contention, does not relinquishes the control over its local medium,
in a particular \textit{sample path}, an mIS of APs (which may not be
an MIS) would \textit{grab} the channel during $T$\footnote{If $W$ is
  large enough, the possibility of two dependent APs transmitting
  together can be ruled out.}. This is not surprising since with
infinite $\rho_i$'s, the CTMC $\{\mathcal{A}(t), t \geq 0\}$ becomes
absorbing with the maximal independent sets of cells as the only
absorbing states and we cannot expect the time average to be equal to
the ensemble average. 

At time $T$, APs that could transmit consecutive broadcast packets
stop contending until time $(M-1) \times T$ and APs that remain
blocked switch to Channel-2, sample fresh back-offs and keep
contending until $2T$ and so on. APs that remain blocked throughout
the duration $(M-1) \times T$ stick to Channel-$M$. Thus, in every
time duration $T$, a maximal independent set of APs would be assigned
a channel. Normal network operation can begin after time $(M-1) \times
T$. Notice that, \textit{mISA does not require any knowledge of AP
  topology and runs in a completely decentralized manner without any
  message passing}. In addition, if there is a central controller to
which the APs can communicate, mISA can be repeated several times
before normal network operation could begin. The central controller,
which obtains the global view of the channel assignments, can choose
the best among the solutions provided by mISA. In absence of
centralized control, mISA can be invoked periodically. Thus, mISA can
be easily implemented in real networks in a completely decentralized
manner if the number of channels for every AP is the same and
known. However, mISA requires loose synchronization among the APs
which necessitates some message passing before mISA could begin.

\section{Conclusions and Future Work}
\label{sec:conclusion}

In this paper, we identified a Pairwise Binary Dependence (PBD)
condition that allows a scalable cell level modeling of WLANs. The PBD
condition is likely to hold at higher PHY rates and denser AP
deployments. We developed a cell level model both under saturation
condition and for TCP-controlled long file downloads. Our analytical
model was shown to be quite accurate, insightful and capable of
comparing few design alternatives. Thus, we believe that our modeling
framework is a significant step toward gaining ``first-cut''
analytical understanding of WLANs with dense deployments of APs. We
demonstrated how our analytical model could be applied along with the
Linear Reward-Inaction learning algorithm for optimal channel
assignment. Based on the insights provided by our model, we also
proposed a simple decentralized algorithm called mISA which can
provide channel assignments that are Nash equilibria in pure
strategies in only as many steps as there are channels. Although mISA
is based on a standard graph coloring technique, its applicability to
the considered problem is strengthen by and its practical
implementability is guided by a rigorous analytical model developed in
this paper. Developing simple and practical algorithms for general
objective functions, other than maximizing the normalized network
throughput, is a topic of our ongoing research.

\bibliographystyle{./IEEEtran}
\bibliography{interferencebib}  

\onecolumn

\appendices

\section{Derivation of Equation \ref{eqn:gamma_i-multicell}} 
\label{app:derivation-eqn-gamma_i}


Consider a tagged node in Cell-$i$. Let \\

\vspace{1mm}

$A_i(T)$ : number of attempts made by the tagged node up to time $T$ 

\vspace{1mm}

$C_i(T)$ : number of collisions as seen by the tagged node up to time $T$

\vspace{1mm}

$A_i^{\mathcal{A}}(T)$ : number of attempts made by the tagged node in
State-${\mathcal{A}}$ up to time $T$ 

\vspace{1mm}

$C_i^{\mathcal{A}}(T)$ : number of collisions as seen by the tagged
node in State-${\mathcal{A}}$ up to time $T$ 

\vspace{1mm}

$B_i^{\mathcal{A}}(T)$ : number of back-off slots elapsed in Cell-$i$
in State-${\mathcal{A}}$ up to time $T$ \\

\noindent Then, we have

\begin{eqnarray}
\label{eqn:gamma_i-multicell-derivation}
\gamma_i &=& \displaystyle \lim_{T \longrightarrow \infty}
\frac{C_i(T)}{A_i(T)} = \frac{\displaystyle \lim_{T \longrightarrow \infty}
  \frac{1}{T} \sum_{\mathcal{A} \in \bmath{\mathcal{A}} \; : \; i \in
    \mathcal{U}_{\mathcal{A}}} C_i^{\mathcal{A}}(T)}{\displaystyle
  \lim_{T \longrightarrow \infty} \frac{1}{T} \sum_{\mathcal{A} \in
    \bmath{\mathcal{A}} \; : \; i \in \mathcal{U}_{\mathcal{A}}}
  A_i^{\mathcal{A}}(T)} = \frac{\displaystyle \lim_{T \longrightarrow \infty}
  \frac{1}{T} \sum_{\mathcal{A} \in \bmath{\mathcal{A}} \; : \; i \in
    \mathcal{U}_{\mathcal{A}}} B_i^{\mathcal{A}}(T) \times
  \frac{C_i^{\mathcal{A}}(T)}{B_i^{\mathcal{A}}(T)}}{\displaystyle
  \lim_{T \longrightarrow \infty} \frac{1}{T} \sum_{\mathcal{A} \in
    \bmath{\mathcal{A}} \; : \; i \in \mathcal{U}_{\mathcal{A}}}
  B_i^{\mathcal{A}}(T) \times
  \frac{A_i^{\mathcal{A}}(T)}{B_i^{\mathcal{A}}(T)}} \nonumber \\ 
&=& \frac{\displaystyle \sum_{\mathcal{A} \in \bmath{\mathcal{A}} \; :
    \; i \in \mathcal{U}_{\mathcal{A}}} \left(\lim_{T \longrightarrow
    \infty} \frac{B_i^{\mathcal{A}}(T)}{T}\right) \times \left(\lim_{T
    \longrightarrow
    \infty}\frac{C_i^{\mathcal{A}}(T)}{B_i^{\mathcal{A}}(T)}\right)}{\displaystyle
  \sum_{\mathcal{A} \in \bmath{\mathcal{A}} \; : \; i \in
    \mathcal{U}_{\mathcal{A}}} \left(\lim_{T \longrightarrow \infty}
  \frac{B_i^{\mathcal{A}}(T)}{T}\right) \times \left(\lim_{T
    \longrightarrow \infty}
  \frac{A_i^{\mathcal{A}}(T)}{B_i^{\mathcal{A}}(T)}\right)} \nonumber
\\ 
&=& \frac{\displaystyle \sum_{\mathcal{A} \in \bmath{\mathcal{A}} \; :
    \; i \in \mathcal{U}_{\mathcal{A}}} \left(\lim_{T \longrightarrow
    \infty} \frac{B_i^{\mathcal{A}}(T)}{T}\right) \times \left(\lim_{T
    \longrightarrow \infty}
  \frac{A_i^{\mathcal{A}}(T)}{B_i^{\mathcal{A}}(T)}\right) \times
  \left(\lim_{T \longrightarrow
    \infty}\frac{C_i^{\mathcal{A}}(T)}{A_i^{\mathcal{A}}(T)}\right)}{\displaystyle
  \sum_{\mathcal{A} \in \bmath{\mathcal{A}} \; : \; i \in
    \mathcal{U}_{\mathcal{A}}} \left(\lim_{T \longrightarrow \infty}
  \frac{B_i^{\mathcal{A}}(T)}{T}\right) \times \left(\lim_{T
    \longrightarrow \infty}
  \frac{A_i^{\mathcal{A}}(T)}{B_i^{\mathcal{A}}(T)}\right)} \nonumber
\\ 
&=& \frac{\displaystyle \sum_{\mathcal{A} \in \bmath{\mathcal{A}} \; :
    \; i \in \mathcal{U}_{\mathcal{A}}}
  \left(\frac{\pi(\mathcal{A})}{\sigma}\right) \times \beta_i \times
  \left(1 - (1-\beta_i)^{n_i-1} \prod_{j \in \mathcal{N}_i \; : \; j
    \in \mathcal{U}_{\mathcal{A}}} (1-\beta_j)^{n_j}
  \right)}{\displaystyle \sum_{\mathcal{A} \in \bmath{\mathcal{A}} \;
    : \; i \in \mathcal{U}_{\mathcal{A}}}
  \left(\frac{\pi(\mathcal{A})}{\sigma}\right) \times \beta_i}
\nonumber \\ 
&=& \frac{\displaystyle \sum_{\mathcal{A} \in \bmath{\mathcal{A}} \; :
    \; i \in \mathcal{U}_{\mathcal{A}}} \pi(\mathcal{A}) \left(1 -
    (1-\beta_i)^{n_i-1} \prod_{j \in \mathcal{N}_i \; : \; j \in
      \mathcal{U}_{\mathcal{A}}} (1-\beta_j)^{n_j}
    \right)}{\displaystyle \sum_{\mathcal{A} \in \bmath{\mathcal{A}}
    \; : \; i \in \mathcal{U}_{\mathcal{A}}} \pi(\mathcal{A})} 
\end{eqnarray}

\vspace{3mm}

\noindent where the last but one step follows from the facts that: (i)
the time spent in State-$\mathcal{A}$ up to time $T$ is equal to
$\pi(\mathcal{A})T$ and if Cell-$i$ is in back-off in
State-$\mathcal{A}$, then $B_i^{\mathcal{A}}(T) =
\frac{\pi(\mathcal{A}) T}{\sigma}$, (ii) the second limit within the
brackets in the numerator is the long-run fraction of back-off slots
in State-$\mathcal{A}$ in which the tagged node attempts which is
equal to $\beta_i$ irrespective of State-$\mathcal{A}$ given that
Cell-$i$ is in back-off in State-$\mathcal{A}$, and (iii) the third
limit within brackets in the numerator is the long-run fraction of
attempts made by the tagged node in State-$\mathcal{A}$ that result in
collisions and depends on the set of neighbors that are also in
back-off in State-$\mathcal{A}$.

\newpage

\section{Proof of Theorem \ref{thm:xi}} 
\label{app:derivation-theorem-xi}

Solving the balance equations for the CTMC $\{\mathcal{A}(t), t \geq
0\}$ (Equation \ref{eqn:stationary-probabilities}) together with the
normalization equation, we obtain 

\begin{equation}
\label{eqn:piPhi-Delta}
\pi(\Phi) = \displaystyle \frac{1}{\displaystyle \sum_{\mathcal{A} \in
    \bmath{\mathcal{A}}} \left(\prod_{i \in \mathcal{A}} \rho_i
  \right)} = \displaystyle \frac{1}{\Delta} \; . 
\end{equation}


Equation \ref{eqn:fraction-of-channel-time} can now be expanded as

\begin{eqnarray}
\label{eqn:xi-expanded}
x_i &=& \sum_{\mathcal{A} \in \bmath{\mathcal{A}} \; : \; i \in
  \mathcal{A}} \pi(\mathcal{A}) \; + \; \sum_{\mathcal{A} \in
  \bmath{\mathcal{A}}\; : \; i \in \mathcal{U}_{\mathcal{A}}}
\pi(\mathcal{A}) \; \; \; \; \; \; \; \; \; \; \; \; \; \; \; \; \; \;
\; \; \; \; \; \; \; \; \; \; \; \; \; \; \; \; \; \; \; \; \; \; \;
\; \; \; \; \; \; \; \; \; \; \; \; \; \; \; \; \; \; \; \; \; \; \;
(\mbox{since $\mathcal{A} \cap \mathcal{U}_{\mathcal{A}} = \Phi$})
\nonumber \\ 
&=& \sum_{\mathcal{A} \in \bmath{\mathcal{A}} \; : \; i \in
  \mathcal{A}} \left( \prod_{j \in \mathcal{A}}\rho_j \right)
\pi(\Phi) \; + \; \sum_{\mathcal{A} \in \bmath{\mathcal{A}}\; : \; i
  \in \mathcal{U}_{\mathcal{A}}} \left( \prod_{j \in
  \mathcal{A}}\rho_j \right) \pi(\Phi) \; \; \; \; \; \; \; \; \; \;
\; \; \; \; \; \; \; \; \; \; \; \; \; \; \; \; \; \; \; \; \; \;
(\mbox{by Equation \ref{eqn:stationary-probabilities}}) \nonumber \\ 
&=& \frac{1}{\Delta} \left[ \sum_{\mathcal{A} \in \bmath{\mathcal{A}}
    \; : \; i \in \mathcal{A}} \left( \prod_{j \in \mathcal{A}}\rho_j
  \right) \; + \; \sum_{\mathcal{A} \in \bmath{\mathcal{A}}\; : \; i
    \in \mathcal{U}_{\mathcal{A}}} \left( \prod_{j \in
    \mathcal{A}}\rho_j \right) \right]\; \; \; \; \; \; \; \; \; \; \;
\; \; \; \; \; \; \; \; \; \; \; \; \; \; \; \; \; \; \; \; \; \; \;
\; \; \; \; \; (\mbox{by Equation \ref{eqn:piPhi-Delta}}) \nonumber \\ 
&=& \frac{1}{\Delta} \left[ \sum_{\mathcal{A} \in \bmath{\mathcal{A}}
    \; : \; i \in \mathcal{A}, \; \mathcal{A} \cap \mathcal{N}_i =
    \Phi} \left( \prod_{j \in \mathcal{A}}\rho_j \right) \; + \;
  \sum_{\mathcal{A} \in \bmath{\mathcal{A}}\; : \; \mathcal{A} \cap
    (\{i\} \cup \mathcal{N}_i) = \Phi} \left( \prod_{j \in
    \mathcal{A}}\rho_j \right) \right] \nonumber \\ 
&=& \frac{1}{\Delta} \left[ \rho_i \sum_{\mathcal{A} \in
    \bmath{\mathcal{A}} \; : \; \mathcal{A} \cap \{i\} = \Phi, \;
    \mathcal{A} \cap \mathcal{N}_i = \Phi} \left( \prod_{j \in
    \mathcal{A}}\rho_j \right) \; + \; \sum_{\mathcal{A} \in
    \bmath{\mathcal{A}}\; : \; \mathcal{A} \cap (\{i\} \cup
    \mathcal{N}_i) = \Phi} \left( \prod_{j \in \mathcal{A}}\rho_j
  \right) \right] \; ,
\end{eqnarray}

\vspace{3mm}

\noindent where the fourth step follows from the facts that: (i) if
Cell-$i$ is active, then none of its neighbors can be active, and (i)
if Cell-$i$ is in back-off, then neither Cell-$i$ nor any of its
neighbors can be active. Observe now that, the independent sets of
$\mathcal{G}_i$ are also independent sets of $\mathcal{G}$. Thus,
$\Delta_i$ can be obtained by restricting the summation in Equation
\ref{eqn:delta} to only those independent sets of $\mathcal{G}$ that
do not contain Cell-$i$ or any cell in $\mathcal{N}_i$, i.e., 

\begin{equation}
\label{eqn:delta_i}
\Delta_i := \displaystyle \sum_{\mathcal{A} \in \bmath{\mathcal{A}} \;
  : \; \mathcal{A} \cap (\{i\} \cup \mathcal{N}_i) = \Phi} \left( \prod_{i
  \in \mathcal{A}}\rho_i \right) = \displaystyle \sum_{\mathcal{A} \in
  \bmath{\mathcal{A}} \; : \; \mathcal{A} \cap \{i\} = \Phi, \;
  \mathcal{A} \cap \mathcal{N}_i = \Phi} \left( \prod_{i \in
  \mathcal{A}}\rho_i \right) \; , 
\end{equation}

\noindent where the second equality follows from the fact that union
of two sets will be equal to the empty set iff each of the sets is an
empty set. Applying Equation \ref{eqn:delta_i} in the last step of
Equation \ref{eqn:xi-expanded}, we obtain Equation
\ref{eqn:xi-in-terms-of-deltai}.

\end{document}